%% file: main.tex
\documentclass[aps, prb, reprint]{revtex4-1}

\usepackage{natbib}
\input{preamble}
\usepackage{algorithm}
\usepackage{algpseudocode}
\usepackage{tabularx}

\DeclareMathOperator{\concat}{concatenate}
\DeclareMathOperator{\matmul}{matmul}
\DeclareMathOperator{\dotproduct}{dot}
\DeclareMathOperator{\envelope}{envelope}

\algnewcommand\algorithmicforeach{\textbf{for each}}
\algdef{S}[FOR]{ForEach}[1]{\algorithmicforeach\ #1\ \algorithmicdo}

\begin{document}

\title{Ab-Initio Solution of the Many-Electron Schr{\"o}dinger Equation with Deep Neural Networks}
\author{David Pfau*}
\author{James S. Spencer*}
\author{Alexander G. D. G. Matthews}
\affiliation{DeepMind, 6 Pancras Square, London N1C 4AG}
\author{W. M. C. Foulkes}
\affiliation{Department of Physics, Imperial College London, South Kensington Campus, London SW7 2AZ}
\date{September 10, 2020}

\begin{abstract}
Given access to accurate solutions of the many-electron Schr{\"o}dinger equation, nearly all chemistry could be derived from first principles. Exact wavefunctions of interesting chemical systems are out of reach because they are NP-hard to compute in general, but approximations can be found using polynomially-scaling algorithms. The key challenge for many of these algorithms is the choice of wavefunction approximation, or Ansatz, which must trade off between efficiency and accuracy. Neural networks have shown impressive power as accurate practical function approximators and promise as a compact wavefunction Ansatz for spin systems, but problems in electronic structure require wavefunctions that obey Fermi-Dirac statistics. Here we introduce a novel deep learning architecture, the Fermionic Neural Network, as a powerful wavefunction Ansatz for many-electron systems. The Fermionic Neural Network is able to achieve accuracy beyond other variational quantum Monte Carlo Ans{\"a}tze on a variety of atoms and small molecules. Using no data other than atomic positions and charges, we predict the dissociation curves of the nitrogen molecule and hydrogen chain, two challenging strongly-correlated systems, to significantly higher accuracy than the coupled cluster method, widely considered the most accurate scalable method for quantum chemistry at equilibrium geometry. This demonstrates that deep neural networks can improve the accuracy of variational quantum Monte Carlo to the point where it outperforms other ab-initio quantum chemistry methods, opening the possibility of accurate direct optimization of wavefunctions for previously intractable many-electron systems.
\end{abstract}

\maketitle

\section{Introduction}

The success of deep learning in artificial intelligence\cite{krizhevsky2012imagenet, silver2016mastering} has led to an outpouring of research into the use of neural networks for quantum physics and chemistry. Many of these methods train a deep neural network to predict properties of novel systems by use of supervised learning on a dataset compiled from existing computational methods --- typically density functional theory (DFT),\cite{gilmer2017neural, schutt2019unifying} exact solutions on a lattice,\cite{mills2017deep} or coupled cluster with single, double and perturbative triple excitations (CCSD(T)).\cite{sinitskiy2019physical,cheng2019universal} Yet all of these methods have drawbacks. Even CCSD(T), which is generally much more accurate than DFT, has difficulties with bond breaking and transition states.\cite{bartlett2007coupled} While methods exist that are even more accurate, most suffer from impractical scaling (in the worst case exponential)\cite{troyer2005complexity} or require complicated system-dependent tuning, making them difficult to apply ``out-of-the-box" to new systems. Here we focus instead on {\em ab-initio} methods that use deep neural networks as approximate solutions to the many-electron Schr{\"o}dinger equation {\em without the need for external data}. We are able to achieve very high accuracy on a number of small but challenging systems, all with the same neural network architecture, suggesting that our method could be a promising ``out-of-the-box" solution for larger systems for which existing approaches are insufficient.

The ground state wavefunction $\psi(\mathbf{x}_1,\mathbf{x}_2,\ldots,\mathbf{x}_n)$ and energy $E$ of a chemical system with $n$ electrons may be found by solving the time-independent Schr{\"o}dinger equation,
\begin{gather}
\hat{H}\psi(\mathbf{x}_1,\ldots,\mathbf{x}_n) = E\psi(\mathbf{x}_1,\ldots,\mathbf{x}_n) \\
\label{eqn:schrodinger}
	\begin{aligned}
		\hat{H} = &-\frac{1}{2}\sum_i \nabla^2_i + \sum_{i > j} \frac{1}{|\mathbf{r}_i-\mathbf{r}_j|} \\
			  &- \sum_{i I} \frac{Z_I}{|\mathbf{r}_i - \mathbf{R}_I|} 	+ \sum_{I > J} \frac{Z_I Z_J}{|\mathbf{R}_I-\mathbf{R}_J|}
	\end{aligned} \nonumber
\end{gather}

\noindent where $\mathbf{x}_i=\{\mathbf{r}_i, \sigma_i\}$ are the coordinates of electron $i$, with $\mathbf{r}_i\in\mathbb{R}^3$ the position and $\sigma_i\in\{\uparrow,\downarrow\}$ the spin, $\nabla^2_i$ is the Laplacian with respect to $\mathbf{r}_i$, and $\mathbf{R}_I$ and $Z_I$ are the position and atomic number of nucleus $I$. We work in the Born-Oppenheimer approximation,\cite{born1927zur} where the nuclear positions are fixed input parameters, and Hartree atomic units are used throughout. The Schr\"{o}dinger differential operator is spin independent but the electron spin matters because the wavefunction must obey Fermi-Dirac statistics --- it must be antisymmetric under the simultaneous exchange of the position and spin coordinates of any two electrons: $\psi(\ldots,\mathbf{x}_i,\ldots,\mathbf{x}_j,\ldots) = -\psi(\ldots,\mathbf{x}_j,\ldots,\mathbf{x}_i,\ldots)$.

Many approaches in quantum chemistry start from a finite set of one-electron orbitals $\phi_1,\ldots,\phi_N$ and approximate the many-electron wavefunction as a linear combination of antisymmetrized tensor products (Slater determinants) of those functions:
\begin{widetext}
\begin{equation}
    \sum_{\mathcal{P}}\mathrm{sign}(\mathcal{P})\prod_i \phi^k_i\left(\mathbf{x}_{\mathcal{P}_i} \right) =
    \left|\begin{matrix}
        \phi^k_1(\mathbf{x}_1) & \ldots & \phi^k_1(\mathbf{x}_n) \\
        \vdots &  & \vdots \\
        \phi^k_n(\mathbf{x}_1) & \ldots & \phi^k_n(\mathbf{x}_n)
    \end{matrix}\right| = \det\left[\phi^k_i(\mathbf{x}_j)\right] =  \det\left[\mathbf{\Phi}^k\right],
    \label{eqn:slater_det}
\end{equation}
\begin{equation}
    \psi(\mathbf{x}_1,\ldots,\mathbf{x}_n) = \sum_k \omega_k 
    \det[\mathbf{\Phi}^k],
    \label{eqn:full_ci_wavefun}
\end{equation}
\end{widetext}
where $\{\phi^k_1,\ldots,\phi^k_n\}$ is a subset of $n$ of the $N$ orbitals, the sum in Eqn.~\ref{eqn:slater_det} is taken over all permutations $\mathcal{P}$ of the electron indices, and the sum in Eqn.~\ref{eqn:full_ci_wavefun} is over all subsets of $n$ orbitals. The difficulty is that the number of possible Slater determinants rises exponentially with the system size, restricting this ``full configuration-interaction'' (FCI) approach to tiny molecules, even with recent advances.\cite{booth2010approaching}

To address problems of practical interest, a more compact representation of the wavefunction is needed. The choice of function class used to approximate the wavefunction is known as the wavefunction Ansatz. For most applications of quantum Monte Carlo (QMC) methods to quantum chemistry, the default choice is the Slater-Jastrow Ansatz,\cite{foulkes2001quantum} which takes a truncated linear combination of Slater determinants and adds a multiplicative term --- the Jastrow factor --- to capture close-range correlations. The Jastrow factor is normally a product of functions of the distances between pairs and triplets of particles. Additionally, a backflow transformation\cite{feynman1956energy} is sometimes applied before the orbitals are evaluated, shifting the position of every electron by an amount dependent on the positions of nearby electrons. There are many alternative Ans\"atze,\cite{bajdich2006pfaffian, orus2014practical} but for continuous-space many-electron problems in three dimensions the Slater-Jastrow-backflow form remains the default.

Here, we greatly improve the accuracy of the Slater-Jastrow-backflow variational quantum Monte Carlo (VMC) method by using a neural network we dub the Fermionic Neural Network, or FermiNet, as a more flexible Ansatz. This avoids the use of a finite basis set, a significant source of error for other Ans{\"a}tze, and models higher-order electron-electron interactions compactly. The use of neural networks as a compact wavefunction Ansatz has been studied before for spin systems\cite{carleo2017solving, choo2018symmetries, nagy2019variational, nomura2017restricted, yang2019deep} and many-electron systems on a lattice\cite{nomura2017restricted, luo2019backflow} as well as small systems of bosons in continuous space.\cite{saito2018method} Applications of neural network Ans{\"a}tze to chemical systems have been limited to date, presumably due to the complexity of Fermi-Dirac statistics. Existing work has been restricted to very small numbers of electrons,\cite{kessler2019artificial} or has been of very low accuracy.\cite{han2018solving} Unlike these other approaches, we use the Slater determinant as the starting point for our Ansatz, and then extend it by generalising the single-electron orbitals to include generic exchangeable nonlinear interactions of {\em all} electrons. In a conventional backflow transformation, the electron positions $\mathbf{r}_j$ at which the one-electron orbitals in the Slater determinants are evaluated are replaced by collective coordinates $\mathbf{r}_j + \sum_{i (\ne j)} \eta(r_{ij})(\mathbf{r}_i-\mathbf{r}_j)$, but the orbitals remain functions of a single three-dimensional variable. The FermiNet wave function goes much further, replacing the one-electron orbitals $\phi_i^k(\mathbf{x}_j)$ by functions of $3n$ independent variables. Every ``orbital'' in every determinant now depends both on $\mathbf{x}_j$ and (in a general symmetric way) on the position and spin coordinates of every other electron.

Our approach is similar in spirit to the neural network backflow transform\cite{luo2019backflow} that has been applied to discrete systems. Certain simplifications in the discrete case allow the use of conventional neural networks, while the continuous case requires a novel architecture to handle antisymmetry constraints, boundary conditions and cusps. The closest prior work we are aware of in continuous space is the iterative backflow transform,\cite{taddei2015iterative, ruggeri2018nonlinear} which has been applied to superfluid $^3$He. While that work uses intermediate layers of the same dimensionality as the input, the FermiNet can use intermediate layers of arbitrary dimensionality, increasing the representational capacity.\footnote{Since this manuscript appeared online, several other works using neural networks as Ans{\"a}tze for continuous-space fermionic systems have appeared.\cite{hermann2019deep, choo2019fermionic} The first\cite{hermann2019deep} also augments the typical Slater-Jastrow Ansatz with a deep neural network, while physical constraints like the cusp conditions are included explicitly. As the model has fewer parameters than ours, it is faster to optimize, but does not achieve the same accuracy. The second\cite{choo2019fermionic} represents a chemical system in second-quantized form using a given basis set, then fits a restricted Boltzmann machine to the ground state. This model is also able to exceed the performance of CCSD(T) {\em within} a basis set. However, it is less clear how easily this model extrapolates to the complete basis set limit. Our approach sidesteps the difficulty of choosing a basis set entirely.}

The FermiNet is not only an improvement over existing Ans{\"a}tze for VMC, but is competitive with and in some cases superior to more sophisticated quantum chemistry algorithms. Projector methods such as diffusion quantum Monte Carlo (DMC)\cite{foulkes2001quantum} and auxiliary field quantum Monte Carlo (AFQMC)\cite{zhang2018abinitio} generate stochastic trajectories that sample the ground state wavefunction without the need for an explicit representation, although accurate explicit trial wavefunctions are still required for good performance and numerical stability. We find the FermiNet is competitive with projector methods on all systems investigated, in contrast with the conventional wisdom that VMC is less accurate. Coupled cluster methods\cite{bartlett2007coupled} use an Ansatz that multiplies a reference wavefunction by an exponential of a truncated sum of creation and annihilation operators. This proves remarkably accurate for equilibrium geometries, but conventional reference wavefunctions are insufficient for systems with many low-lying excited states. We evaluate the FermiNet on a variety of stretched systems and find that it outperforms coupled cluster in all cases.

\begin{figure*}
    \centering
    \includegraphics[width=\textwidth]{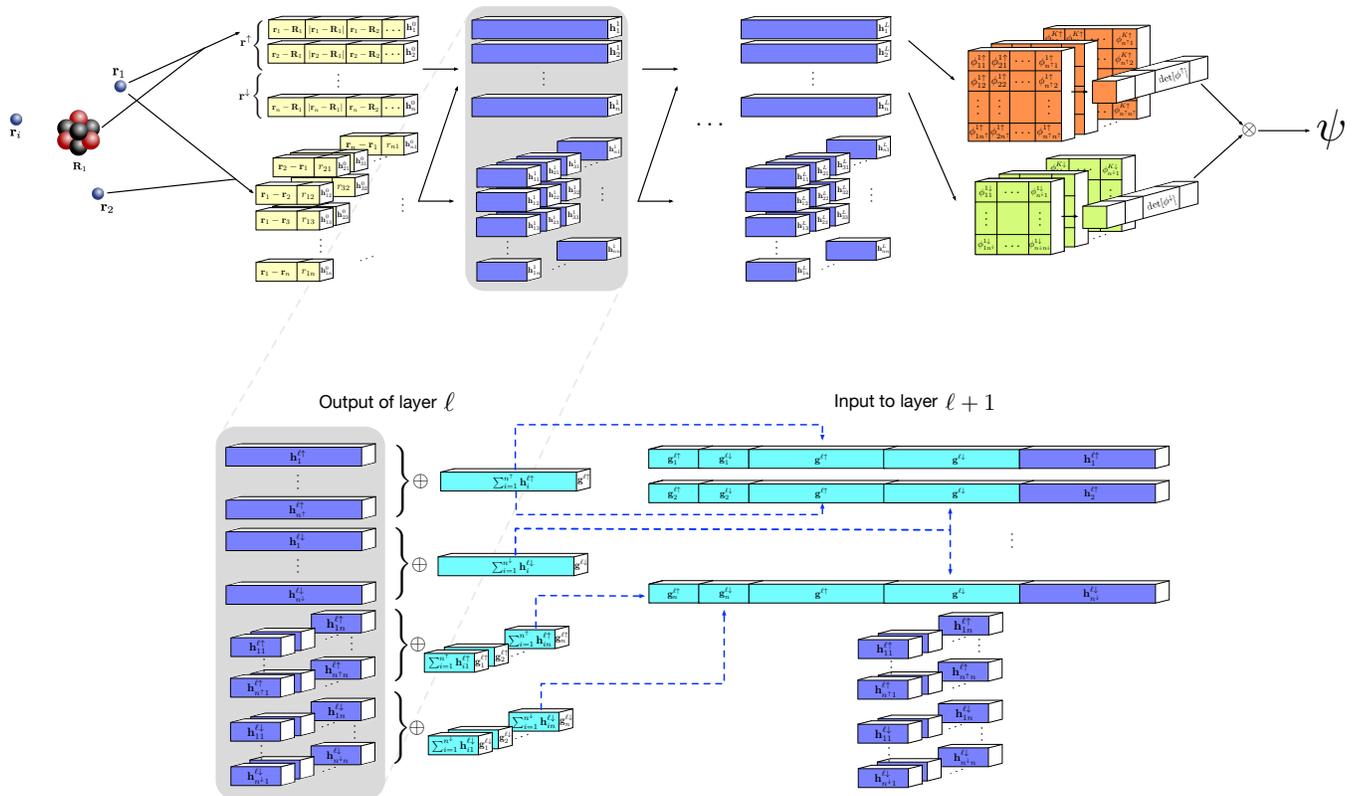}\\
    \caption{The Fermionic Neural Network (FermiNet). Top: Global architecture. Features of one or two electron positions are inputs to different streams of the network. These features are transformed through several layers, a determinant is applied, and the wavefunction at that position is given as output. Bottom: Detail of a single layer. The network averages features of electrons with the same spin together, then concatenates these features to construct an equivariant function of electron position at each layer.}
    \label{fig:architecture}
\end{figure*}

\section{Fermionic Neural Networks}

\subsection{Fermionic Neural Network Architecture}

\begin{figure}[t]
\begin{algorithm}[H]
   \begin{algorithmic}[1]
       \Require walker configuration $\{\mathbf{r}_1^\uparrow, \cdots, \mathbf{r}_{n^\uparrow}^\uparrow, \mathbf{r}_1^\downarrow, \cdots, \mathbf{r}_{n^\downarrow}^\downarrow\}$
       \Require nuclear positions $\{\mathbf{R}_I\}$
       \ForEach {electron $i,\alpha$}
           \State $\mathbf{h}_{i}^{\ell \alpha} \gets \concat(\mathbf{r}^\alpha_i - \mathbf{R}_I, |\mathbf{r}^\alpha_i - \mathbf{R}_I|\ \forall\ I)$ 
           \State $\mathbf{h}_{ij}^{\ell \alpha\beta} \gets \concat(\mathbf{r}^\alpha_i - \mathbf{r}^\beta_j, |\mathbf{r}^\alpha_i - \mathbf{r}^\beta_j|\ \forall\ j,\beta)$ 
       \EndFor
       \ForEach {layer $\ell \in \{0, L-1\}$}
           \State $\mathbf{g}^{\ell \uparrow} \gets \frac{1}{n^{\uparrow}}\sum_{i}^{n^{\uparrow}} \mathbf{h}_{i}^{\ell \uparrow}$
           \State $\mathbf{g}^{\ell \downarrow} \gets \frac{1}{n^{\downarrow}}\sum_{i}^{n^{\downarrow}} \mathbf{h}_{i}^{\ell \downarrow}$
           \ForEach {electron $i,\alpha$}
               \State $\mathbf{g}_i^{\ell \alpha\uparrow} \gets \frac{1}{n^{\uparrow}}\sum_j^{n^\uparrow} \mathbf{h}_{ij}^{\ell \alpha\uparrow}$
                \State $\mathbf{g}_i^{\ell \alpha\downarrow} \gets \frac{1}{n^{\downarrow}}\sum_j^{n^\downarrow} \mathbf{h}_{ij}^{\ell \alpha\downarrow}$
               \State $\mathbf{f}_i^{\ell \alpha} \gets \concat(\mathbf{h}_i^{\ell \alpha}, \mathbf{g}^{\ell \uparrow}, \mathbf{g}^{\ell \downarrow}, \mathbf{g}_i^{\ell \alpha\uparrow}, \mathbf{g}_i^{\ell \alpha\downarrow})$
               \State $\mathbf{h}_i^{\ell+1 \alpha} \gets \tanh\left( \matmul(\mathbf{V}^l, \mathbf{f}_i^{\ell \alpha}) + \mathbf{b}^l \right) + \mathbf{h}_i^{\ell \alpha}$
               \State $\mathbf{h}_{ij}^{\ell+1 \alpha\beta} \gets \tanh\left( \matmul(\mathbf{W}^l, \mathbf{h}_{ij}^{\ell \alpha\beta}) + \mathbf{c}^l \right) + \mathbf{h}_{ij}^{\ell \alpha\beta}$
           \EndFor
       \EndFor
       \ForEach {determinant $k$}
           \ForEach {orbital $i$}
               \ForEach {electron $j,\alpha$}
                   \State $e \gets \envelope(\mathbf{r}_j^\alpha, \{r_i^\alpha - \mathbf{R}_I\})$
                   \State $\phi_i(\mathbf{r}_j^\alpha; \{\mathbf{r}_{/j}^{\alpha}\}; \{\mathbf{r}^{\bar{\alpha}}\}) = \left(\dotproduct(\mathbf{w}^{k\alpha}_i, \mathbf{h}^{L\alpha}_j) + g^{k\alpha}_i\right) e$
               \EndFor
           \EndFor
           \State $D^{k\uparrow} \gets \det\left[\phi^{k \uparrow}_i(\mathbf{r}^\uparrow_j; \{\mathbf{r}^\uparrow_{/j}\}; \{\mathbf{r}^\downarrow\})\right]$
           \State $D^{k\downarrow} \gets \det\left[\phi^{k \downarrow}_i(\mathbf{r}^\downarrow_j; \{\mathbf{r}^\downarrow_{/j}\}; \{\mathbf{r}^\uparrow\})\right]$           
       \EndFor
       \State $\psi \gets \sum_k \omega_k D^{k\uparrow} D^{k\downarrow}$
   \end{algorithmic}
  \caption{FermiNet evaluation.}
  \label{algo:ferminet} 
\end{algorithm}
\end{figure}

To construct an expressive neural network Ansatz, we note that nothing requires the orbitals in the matrix in Eqn.~\ref{eqn:slater_det} to be functions of the coordinates of a single electron. The only requirement for the determinant of a matrix-valued function of $\mathbf{x}_1$, $\mathbf{x}_2$, $\ldots$, $\mathbf{x}_n$ to be antisymmetric is that exchanging any two input variables, $\mathbf{x}_i$ and $\mathbf{x}_j$, exchanges two rows or columns of the output matrix, leaving the rest invariant. This observation allows us to replace the single-electron orbitals $\phi^k_i(\mathbf{x}_j)$  by multi-electron functions $\phi^k_i(\mathbf{x}_j; \mathbf{x}_1,\ldots,\mathbf{x}_{j-1},\mathbf{x}_{j+1},\ldots,\mathbf{x}_n) = \phi^k_i(\mathbf{x}_j; \{\mathbf{x}_{/j}\})$, where $ \{\mathbf{x}_{/j}\}$ denotes the set of all electron states except $\mathbf{x}_j$, so long as these functions are invariant to any change in the order of the arguments after $\mathbf{x}_j$. In theory, a single determinant made up of these permutation-equivariant functions is sufficient to represent any antisymmetric function (see Appendix~\ref{sec:hutter}), however the practicality of approximating an antisymmetric function will depend on the choice of permutation-equivariant function class; we hence use a small linear combination of $n_k$ determinants in this work. The construction of a set of these permutation-equivariant functions with a neural network is the main innovation of the FermiNet. We emphasize that determinants constructed from permutation-equivariant functions are substantially more expressive than conventional Slater determinants. Fig.~\ref{fig:architecture} contains a schematic of the network and Algorithm~\ref{algo:ferminet} pseudocode for evaluating the network. 

The Fermionic Neural Network takes features of single electrons and pairs of electrons as input. As input to the single-electron stream of the network, we include both the difference in position between each electron and nucleus $\mathbf{r}_i - \mathbf{R}_I$ and the distance $|\mathbf{r}_i - \mathbf{R}_I|$. The input to the two-electron stream is similarly the differences $\mathbf{r}_i - \mathbf{r}_j$ and distances $|\mathbf{r}_i - \mathbf{r}_j|$. Adding the absolute distances between particles directly as input removes the need to include a separate Jastrow factor after a determinant. As the distance is a non-smooth function at zero, the neural network is capable of expressing the non-smooth behavior of the wavefunction when two particles coincide --- the wavefunction cusps. Accurately modeling the cusps is critical for correctly estimating the energy and other properties of the system. The quality of the wavefunction cusps for the helium atom are investigated in Appendix~\ref{sec:cusps}. We denote the concatenation of all features for one electron $\mathbf{h}^0_i$, or $\mathbf{h}^{0\alpha }_i$ if we explicitly index its spin $\alpha\in\{\uparrow,\downarrow\}$; the features of two electrons are denoted $\mathbf{h}^0_{ij}$ or $\mathbf{h}^{0\alpha\beta}_{ij}$. If the system has $n^\uparrow$ spin-up electrons and $n^\downarrow$ spin down electrons, without loss of generality we can reorder the electrons so that $\sigma_j = \uparrow$ for $j\in 1,\ldots, n^\uparrow$ and $\sigma_j=\downarrow$ for $j\in n^\uparrow + 1, \ldots, n$.

To satisfy the overall antisymmetry constraint for a fermionic wavefunction, intermediate layers of the Fermionic Neural Network must mix information together in a permutation-equivariant way. Permutation-equivariant neural network layers like self-attention have gained success in recent years in natural language processing\cite{vaswani2017attention} and protein folding,\cite{senior2020improved} but we pursue a simpler yet effective approach. Permutation-equivariant layers have also been widely adopted in the computational chemistry and machine learning community for modeling energies and force fields from atomic configurations. \cite{schutt2017quantum, gilmer2017neural, schutt2018schnet} The Fermionic Neural Network shares some architectural details with these models, such as the use of pairwise distances as inputs and parallel streams of feature vectors, one per particle, through the network, but is tailored specifically for mapping electronic configurations to wavefunction values with fixed atomic positions, rather than mapping atomic positions to total energies and other properties.

In our intermediate layers, we take the mean of activations from different streams of the network, concatenate these mean activations together and append them to the single-electron streams of the network. For a single layer this is a purely linear operation, but when combined with a nonlinear activation function after each layer it becomes a flexible architecture for building permutation-equivariant functions\cite{shawetaylor1989building}. Information from both the other one-electron streams and the two-electron streams are fed into the one-electron streams. However, to reduce the computational overhead, no information is transferred between two-electron streams --- these are multilayer perceptrons running in parallel. If the outputs of the one-electron stream at layer $\ell$ with spin $\alpha$ are denoted $\mathbf{h}^{\ell \alpha}_i$ and outputs of the two-electron stream are $\mathbf{h}^{\ell \alpha\beta}_{ij}$, then the input to the one-electron stream for electron $i$ with spin $\alpha$ at layer $\ell+1$ is
\begin{align}
    &\left(
    \mathbf{h}^{\ell\alpha}_i,
    \frac{1}{n^\uparrow}\sum_{j=1}^{n^\uparrow} \mathbf{h}^{\ell\uparrow}_j, \frac{1}{n^\downarrow} \sum_{j=1}^{n^\downarrow} \mathbf{h}^{\ell\downarrow}_j,
    \frac{1}{n^\uparrow} \sum_{j=1}^{n^\uparrow} \mathbf{h}^{\ell\alpha\uparrow}_{ij},
    \frac{1}{n^\downarrow} \sum_{j=1}^{n^\downarrow} \mathbf{h}^{\ell\alpha\downarrow}_{ij}\right) \nonumber \\
    &\qquad =
    \left(\mathbf{h}^{\ell\alpha}_i, \mathbf{g}^{\ell\uparrow}, \mathbf{g}^{\ell\downarrow}, \mathbf{g}^{\ell\alpha\uparrow}_i, \mathbf{g}^{\ell\alpha\downarrow}_i\right) = \mathbf{f}^{\ell \alpha}_i,
    \label{eqn:ferminet_layer}
\end{align}
which is the concatenation of the mean activation for spin up and down parts of the one and two electron streams, respectively. This concatenated vector is then input into a linear layer followed by a tanh nonlinearity. A residual connection is also added between layers of the same shape, for both one and two electron streams:

\begin{figure*}[t]
    \centering
        \begin{minipage}[b]{0.48\textwidth}
            \includegraphics[width=\textwidth]{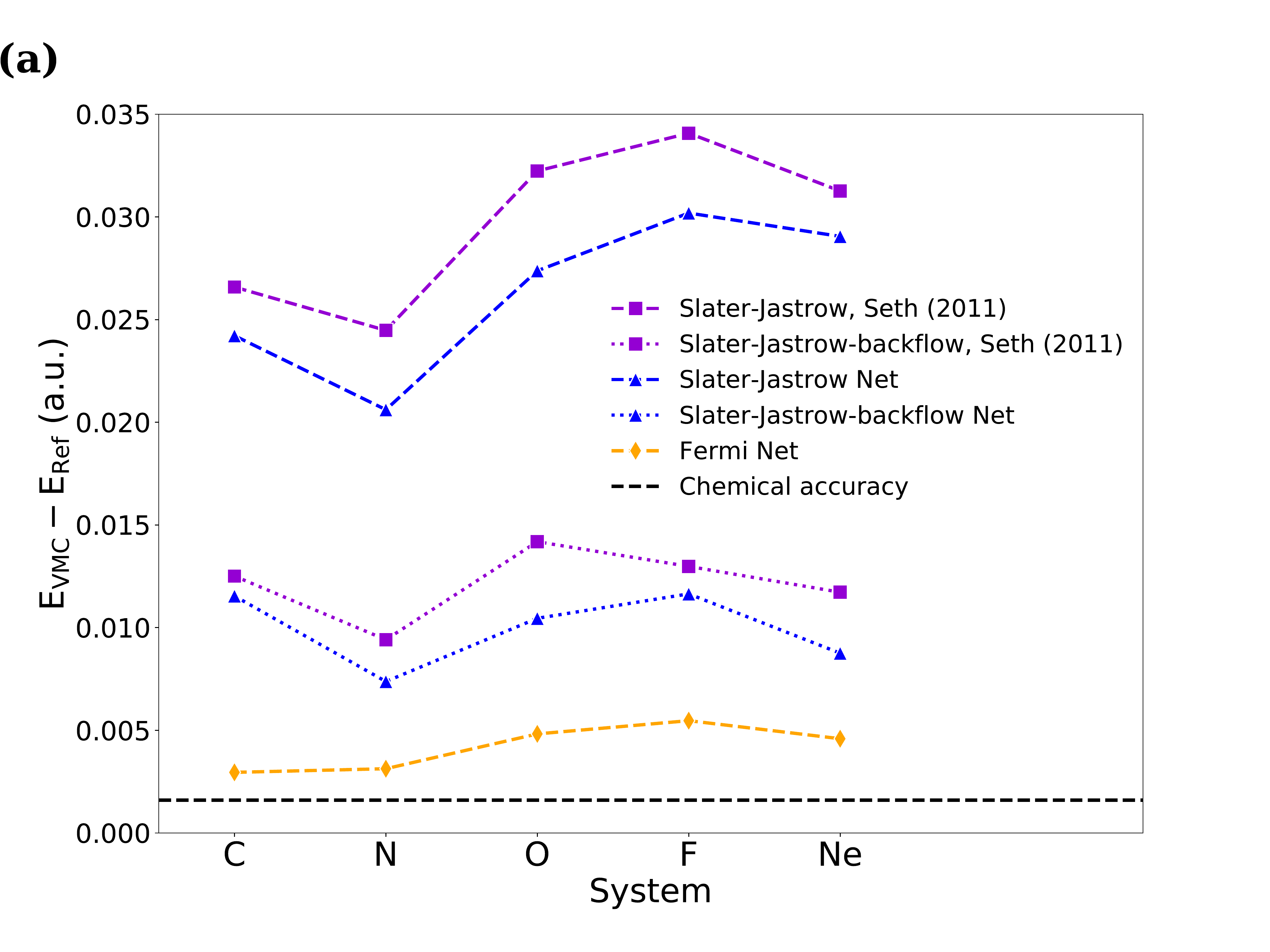}
        \end{minipage}
        \begin{minipage}[b]{0.48\textwidth}
            \includegraphics[width=\textwidth]{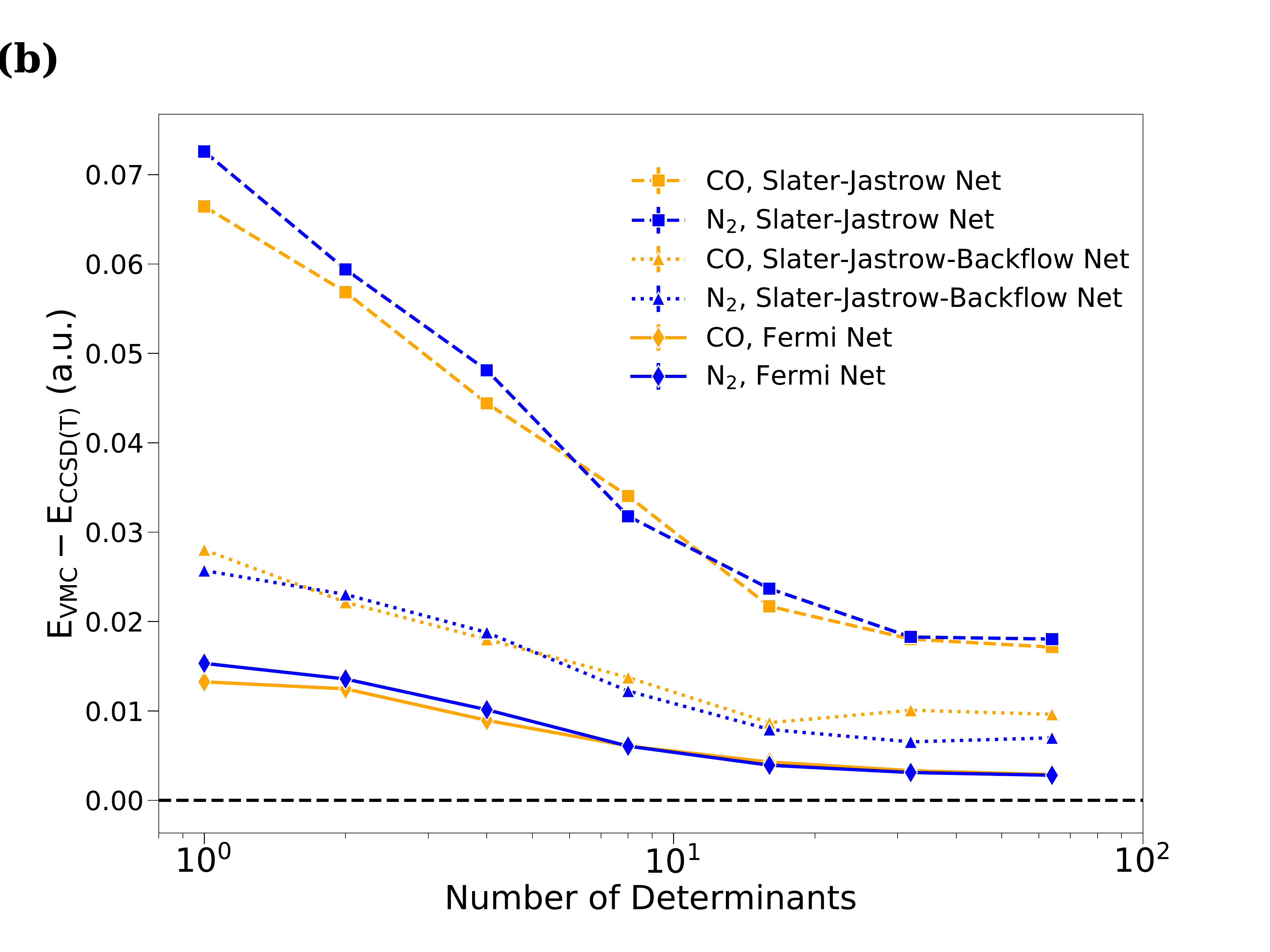}
        \end{minipage}
    \vspace{0.01cm}
    \caption{Comparison of the FermiNet against the Slater-Jastrow Ansatz, with and without backflow. \textbf{(a)}: First-row atoms with a single determinant. Baseline numbers are from Chakravorty \emph{et al.}~\cite{chakravorty1993ground}. The Slater-Jastrow neural network yields slightly lower energies than VMC with a conventional Slater-Jastrow Ansatz, while the FermiNet is substantially more accurate. \textbf{(b)}: The CO and N$_2$ molecules (bond lengths 2.17328 $a_0$ and 2.13534 $a_0$ respectively) with increasing numbers of determinants. All-electron CCSD(T)/CBS results are used as the baseline. No matter how many determinants are used, the FermiNet far exceeds the accuracy of the Slater-Jastrow net.}
    \label{fig:compare_det}
\end{figure*}

\begin{align}
    \mathbf{h}^{\ell+1 \alpha}_i &= \mathrm{tanh}\left(\mathbf{V}^\ell \mathbf{f}^{\ell \alpha}_i + \mathbf{b}^\ell\right) + \mathbf{h}^{\ell\alpha}_i \nonumber \\
    \mathbf{h}^{\ell+1 \alpha\beta}_{ij} &= \mathrm{tanh}\left(\mathbf{W}^\ell\mathbf{h}^{\ell \alpha\beta}_{ij} + \mathbf{c}^\ell\right) + \mathbf{h}^{\ell \alpha\beta}_{ij}
\end{align}

After the last intermediate layer of the network, a final spin-dependent linear transformation is applied to the activations, and the output is multiplied by a weighted sum of exponentially-decaying envelopes, which enforces the boundary condition that the wavefunction goes to zero far away from the nuclei:
\begin{multline}
    \phi^{k\alpha}_i(\mathbf{r}^\alpha_j; \{\mathbf{r}^\alpha_{/j}\}; \{\mathbf{r}^{\bar{\alpha}}\}) =
    \left(\mathbf{w}^{k\alpha}_i \cdot \mathbf{h}^{L\alpha}_j + g^{k\alpha}_i\right)\\
	\sum_{m} \pi^{k\alpha}_{im}\mathrm{exp}\left(-|\mathbf{\Sigma}_{im}^{k \alpha}(\mathbf{r}^{\alpha}_j-\mathbf{R}_m)|\right),
\end{multline}
where $\bar{\alpha}$ is $\downarrow$ if $\alpha$ is $\uparrow$ or vice versa, $\mathbf{h}^{L\alpha}_j$ is an output from the $L$-th  (final) layer of the intermediate single-electron features network for electrons of spin $\alpha$, and $\mathbf{w}^{k\alpha}_i$ ($g^{k\alpha}_i)$ are the weights (biases) of the final linear transformation for determinant $k$. The learned parameters $\pi^{k\alpha}_{im}$ and $\mathbf{\Sigma}_{im}^{k\alpha}\in\mathbb{R}^{3\times3}$ control the anisotropic decay to zero far from each nucleus.
The functions $\{\phi^{k\alpha}_i(\mathbf{r}^\alpha_j)\}$ are then used as the input to multiple determinants, and the full wavefunction is taken as a weighted sum of these determinants:
\begin{align}
	\psi(\mathbf{r}^\uparrow_1,\ldots,\mathbf{r}^\downarrow_{n^\downarrow}) = \sum_{k}\omega_k &\left(\det\left[\phi^{k \uparrow}_i(\mathbf{r}^\uparrow_j; \{\mathbf{r}^\uparrow_{/j}\}; \{\mathbf{r}^\downarrow\})\right]\right.\nonumber\\
	&\left.\hphantom{\left(\right.}\det\left[\phi^{k\downarrow}_i(\mathbf{r}^\downarrow_j; \{\mathbf{r}^\downarrow_{/j}\});
	\{\mathbf{r}^\uparrow\};\right]\right).
    \label{eqn:wavefunction}
\end{align}
Eq.~\ref{eqn:wavefunction} uses the fact that the full determinant $\det[\mathbf{\Phi}] = \det\left[\phi_i(\mathbf{x}_j; \{\mathbf{x}_{/j}\})\right]$ may be replaced by a product of spin-up and spin-down terms if we choose $\phi_i(\mathbf{x}_{j}; \{\mathbf{x}_{/j}\}) = 0$ if $i\in 1\ldots n^\uparrow$ and $j\in n^\uparrow + 1, \ldots, n$ or $i\in n^\uparrow + 1,\ldots, n$ and $j \in 1,\ldots,n^\uparrow$. Then the matrix $\mathbf{\Phi}$ is block-diagonal and:
%
\begin{align}
&\det\left[\mathbf{\Phi}\right] = \det\left[\phi_i(\mathbf{x}_j;\{\mathbf{x}_{/j}\})\right] = \nonumber \\
&\det\left[\phi^\uparrow_i(\mathbf{r}^\uparrow_j;\{\mathbf{r}^\uparrow_{/j}\}; \{\mathbf{r}^\downarrow\})\right]\det\left[\phi^\downarrow_i(\mathbf{r}^\downarrow_j; \{\mathbf{r}^\downarrow_{/j}\}; \{\mathbf{r}^\uparrow\})\right] 
\label{eqn:dets}
\end{align}
%
The new wavefunction is only antisymmetric under exchange of electrons of the same spin, $\{\mathbf{r}^\uparrow\}$ or $\{\mathbf{r}^\downarrow\}$, but nevertheless yields correct expectation values of spin-independent observables and the fully antisymmetric wavefunction can be reconstructed if required. This factorization allows spin-dependence to be handled explictly rather than as input to the network.

The linear combination of determinants in Eq.~\ref{eqn:wavefunction} bears some resemblance to Ans{\"a}tze used in truncated configuration interaction methods like CI singles and doubles (CISD), which are known to have issues with size-consistency, thus it is natural to wonder if the FermiNet also has these issues. The determinants in the FermiNet are very different from conventional Slater determinants, as they allow for essentially arbitrary correlations between electrons in each orbital $\phi^{k\alpha}_i$. We prove in Appendix~\ref{sec:hutter} that a single determinant of this form is in theory general enough to represent {\em any} antisymmetric function, though in practice we require a small number of determinants to reach high accuracy. This may be due to the limitations of finite-size neural networks in representing functions of the type described in Appendix~\ref{sec:hutter}. In all our experiments on N$_2$ and the hydrogen chain (Secs.~\ref{sec:nitrogen_dimer},~\ref{sec:hydrogen_chain}, Table~\ref{tab:hn_sep}), the FermiNet was able to learn a size-consistent solution.

\subsection{Wavefunction optimization}

As in the standard setting for wavefunction optimization for variational Monte Carlo, we sought to minimize the energy expectation value of the wavefunction Ansatz:
\[
\mathcal{L}(\theta) = \frac{\langle \psi_\theta | \hat{H} | \psi_\theta \rangle}{\langle \psi_\theta | \psi_{\theta} \rangle} = \frac{\int d\mathbf{X} \, \psi_\theta^{\ast}(\mathbf{X})\hat{H}\psi_\theta(\mathbf{X})}{\int d\mathbf{X} \, \psi_\theta^{\ast}(\mathbf{X}) \psi_{\theta}(\mathbf{X})} ,
\]
where $\theta$ are the parameters of the Ansatz, $\hat{H}$ is the Hamiltonian of the system as given in Eqn.~\ref{eqn:schrodinger}, and $\mathbf{X}=(\mathbf{x}_1,\ldots,\mathbf{x}_n)$ denotes the full state of all electrons. As $\hat{H}$ is time-reversal invariant and Hermitian, its eigenfunctions and eigenvalues are real. If the minimization is taken over all real normalizable functions, the minimum of the energy occurs when $\psi_{\theta}(\mathbf{X})$ is the ground-state eigenfunction of $\hat{H}$; for a more restricted Ansatz, the minimum lies above the ground-state eigenvalue. When samples from the probability distribution defined by the wavefunction Ansatz $p(\mathbf{X}) \propto \psi^2_\theta(\mathbf{X})$ are available, unbiased estimates of the gradient of the energy with respect to $\theta$ can be computed as follows:
\begin{align}
    \localEnergy(\mathbf{X}) &= \psi^{-1}(\mathbf{X})\hat{H}\psi(\mathbf{X}) , \nonumber \\
    \nabla_\theta \mathcal{L} &= 2\expectp{\left(\localEnergy-\expectp{\localEnergy}\right)\nabla_\theta \mathrm{log}|\waveFunc|} ,
    \label{eqn:vmc_grad}
\end{align}
where $E_L$ is the {\em local energy} and we have dropped the dependence of $\psi$ on $\theta$ for clarity. Recent developments,\cite{clark_computing_2011,neuscamman_optimizing_2012,assaraf_optimizing_2017,zhang2018abinitio} 
including investigating first-order stochastic opitimization methods from the machine learning community.\cite{otis2019complementary,sabzevari_accelerated_2020}, have enabled optimization of conventional wavefunctions with large parameter sets. We use a second-order method which can exploit the structure of the neural network. 

For all wavefunction Ans\"atze used in this paper, the determinants were computed in the log domain, and the final network output gave the log of the absolute value of the wavefunction, along with its sign. The local energy was computed directly in the log domain using the formula:
\begin{align*}
\localEnergy(\overallState) &= \waveFunc^{-1}(\overallState)\hamiltonian\waveFunc(\overallState) \\
&= -\frac{1}{2}\sum_i \left [ \frac{\partial^2 \mathrm{log}|\waveFunc|}{\partial r_i^2}\bigg\rvert_{\mathbf{X}}  +  \left(\frac{\partial \mathrm{log}|\waveFunc|}{\partial r_i}\bigg\rvert_{\mathbf{X}}\right)^{2} \right ] + V(\overallState) ,
\end{align*}
where $V(\overallState)$ is the potential energy of the state $\overallState$ and the index $i$ runs over all 3N dimensions of the electron position vector.
To optimize the wavefunction, we used a modified version of Kronecker Factorized Approximate Curvature (KFAC),\cite{martens2015optimizing} an approximation to natural gradient descent\cite{amari1998natural} appropriate for neural networks. Natural gradient descent updates for optimizing $\mathcal{L}$ with respect to parameters $\theta$ have the form $\delta\theta \propto \mathcal{F}^{-1}\nabla_\theta \mathcal{L}(\theta)$, where $\mathcal{F}$ is the Fisher Information Matrix (FIM):

\[
\mathcal{F}_{ij} = \expectp{\frac{\partial \mathrm{log}p(\overallState)}{\partial \theta_i}  \frac{\partial \mathrm{log}p(\overallState)}{\partial \theta_j}}.
\]
This is equivalent to stochastic reconfiguration\cite{sorella1998green} when the probability density is unnormalized (see Appendix~\ref{sec:ngd_and_sr}) and closely related to the linear method of Toulouse and Umrigar.\cite{toulouse2007optimization}

\begingroup
\squeezetable
\begin{table*}[t]
	\centering
	\hspace*{-1cm}\begin{tabular}{cccccccccccccc}\hline\hline 
        & \multicolumn{7}{c}{Ground state energy (E$_h$)} & \multicolumn{3}{c}{Ionization potential (mE$_h$)}  & \multicolumn{3}{c}{Electron affinity (mE$_h$)}   \\
        \cmidrule(lr){2-8}\cmidrule(lr){9-11}\cmidrule(lr){12-14}
        Atom & FermiNet & VMC\cite{seth2011quantum} & DMC\cite{seth2011quantum} & CCSD(T)/CBS & HF/CBS & Exact\cite{chakravorty1993ground} & \% corr & FermiNet & Expt.\cite{klopper2010submev} & $\Delta$E & FermiNet & Expt.\cite{klopper2010submev} & $\Delta$E \\\hline
        Li & -7.47798(1)        & -7.478034(8) & {\bf -7.478067(5)} & -7.478157 & -7.432747 & -7.47806032 & 99.82(3) & 198.10(4)  & 198.147 & 0.04(4)   & 21.82(20)  & 22.716  & 0.89(20) \\
        Be & {\bf -14.66733(3)} & -14.66719(1) & -14.667306(7)      & -14.66737 & -14.57301 & -14.66736   & 99.97(3) & 342.77(18) & 342.593 & -0.17(18) & -          & -       & -        \\  
        B  & -24.65370(3)       & -24.65337(4) & {\bf -24.65379(3)} & -24.65373 & -24.53316 & -24.65391   & 99.83(3) & 304.86(4)  & 304.979 & 0.12(4)   & 9.03(11)   & 10.336  & 1.31(11) \\
        C  & {\bf -37.84471(5)} & -37.84377(7) & -37.84446(6)       & -37.8448  & -37.6938  & -37.8450    & 99.81(3) & 413.98(8)  & 414.014 & 0.03(8)   & 46.18(9)   & 46.610  & 0.43(9) \\
        N  & {\bf -54.58882(6)} & -54.5873(1)  & -54.58867(8)       & -54.5894  & -54.4047  & -54.5892    & 99.80(3) & 534.80(12) & 534.777 & -0.03(12) & -          & -       & -        \\  
        O  & {\bf -75.06655(7)} & -75.0632(2)  & -75.0654(1)        & -75.0678  & -74.8192  & -75.0673    & 99.70(3) & 500.29(26) & 500.453 & 0.17(26)  & 53.55(19)  & 53.993  & 0.44(19) \\
        F  & {\bf -99.7329(1)}  & -99.7287(2)  & -99.7318(1)        & -99.7348  & -99.4168  & -99.7339    & 99.69(3) & 640.86(41) & 640.949 & 0.09(41)  & 125.71(26) & 125.959 & 0.25(26) \\
        Ne & {\bf -128.9366(1)} & -128.9347(2) & {\bf -128.9366(1)} & -128.9394 & -128.5479 & -128.9376   & 99.74(3) & 794.30(52) & 794.409 & 0.11(52)  & -          & -       & -        \\  
       \hline\hline 
    \end{tabular}
    \caption{Ground state energy, ionization potential and electron affinity for first-row atoms. The QMC method (FermiNet, conventional VMC or DMC) closest to the exact ground state energy for each atom is in bold. Electron affinities for Be, N and Ne are not computed as their anions are unstable. Experimental ionization potentials and electron affinities have had estimated relativistic effects\cite{klopper2010submev} removed. All ground state energies are within chemical accuracy of the exact numerical solution, and all electron affinities and all ionization potentials except neon are within chemical accuracy of experimental results. If no citation is provided, the number was from our own calculation.}
    \label{tab:energy}
\end{table*}
\endgroup

\begin{table*}[t]
\centering
\begin{tabular}{ccccccccc}\hline\hline
        Molecule      & Bond length (a$_0$) & FermiNet (E$_h$)   & \multicolumn{3}{c}{CCSD(T) (E$_h$)}         & HF (E$_h$)     & Exact (E$_h$)                                   & \% corr \\ \cline{4-6}
                      &                     &                     & aug-cc-pCVQZ & aug-cc-pCV5Z & CBS           & CBS             &                                                 & \\\hline
        LiH           & 3.015               & -8.07050(1)         &  -8.0687     &  -8.0697     & -8.070696     & -7.98737        & -8.070548\cite{cencek2000benchmark}             & 99.94(1) \\
        Li$_2$        & 5.051               & -14.99475(1)        &  -14.9921    &  -14.9936    & -14.99507     & -14.87155       & -14.9954\cite{filippi1996multiconfiguration}    & 99.47(1) \\
        NH$_3$        & -                   & -56.56295(8)        &  -56.5535    &  -56.5591    & -56.5644      & -56.2247        & -                                               & 99.57(2) \\
        CH$_4$        & -                   & -40.51400(7)        &  -40.5067    &  -40.5110    & -40.5150      & -40.2171        & -                                               & 99.66(3) \\
        CO            & 2.173               & -113.3218(1)        &  -113.3047   &  -113.3154   & -113.3255     & -112.7871       & -                                               & 99.32(3) \\
        N$_2$         & 2.068               & -109.5388(1)        &  -109.5224   &  -109.5327   & -109.5425     & -108.9940       & -109.5423\cite{filippi1996multiconfiguration}   & 99.36(2) \\
        Ethene        & -                   & -78.5844(1)         &  -78.5733    &  -78.5812    & -78.5888      & -78.0705        & -                                              & 99.16(2) \\
        Methylamine   & -                   & -95.8554(2)         &  -95.8437    &  -           & -95.8653      & -95.2628        & -                                               & 98.36(3) \\
        Ozone         & -                   & -225.4145(3)        &  -225.3907   &  -225.4119   & -225.4338     & -224.3526       & -                                               & 98.42(3) \\
        Ethanol       & -                   & -155.0308(3)        &  -155.0205   &  -           & -155.0545     & -154.1573       & -                                               & 97.36(4) \\
        Bicyclobutane & -                   & -155.9263(6)    &  -155.9216   &  -           & -155.9575     & -154.9372       & -                                               & 96.94(5) \\ \hline\hline
    \end{tabular}
    \caption{Ground state energy at equilibrium geometry for diatomics and small molecules. The percentage of correlation energy captured by the FermiNet relative to the exact energy (where available) or CCSD(T)/CBS is given in the rightmost column. If no citation is provided, the number was from our own calculation. Geometries for larger molecules are given in Appendix~\ref{sec:geometries}.}
    \label{tab:energy_molecules}
\end{table*}

For large neural networks with thousands to millions of parameters, solving the linear system $\mathcal{F}\delta\theta = \nabla_\theta \mathcal{L}$ becomes impractical. KFAC ameliorates this with two approximations. First, any terms $\mathcal{F}_{ij}$ are assumed to be zero when $\theta_i$ and $\theta_j$ are in different layers of the network. This makes the FIM block-diagonal and significantly more efficient to invert. The second approximation is based on the structure of the gradient for a linear layer in a neural network. If $W_\ell$ is the weight matrix for layer $\ell$ of a network, then the block of the FIM for that weight is, in vectorized form:
\begin{multline}
\expectp{\frac{\partial \mathrm{log}p(\overallState)}{\partial \mathrm{vec}(\mathbf{W}_\ell)}  \frac{\partial \mathrm{log}p(\overallState)}{\partial \mathrm{vec}(\mathbf{W}_\ell)}^T} =\\ \expectp{\left(\mathbf{a}_\ell\otimes\mathbf{e}_\ell\right)\left(\mathbf{a}_\ell\otimes\mathbf{e}_\ell\right)^T}
\end{multline}
where $\mathbf{a}_\ell$ are the forward activations and $\mathbf{e}_\ell$ are the backward sensitivities for that layer. KFAC approximates the inverse of this block as the Kronecker product of the inverse second moments:
\begin{multline}
\expectp{\left(\mathbf{a}_\ell\otimes\mathbf{e}_\ell\right)\left(\mathbf{a}_\ell\otimes\mathbf{e}_\ell\right)^T}^{-1} \approx \\ \expectp{\mathbf{a}_\ell\mathbf{a}_\ell^T}^{-1}\otimes\expectp{\mathbf{e}_\ell\mathbf{e}_\ell^T}^{-1}
\end{multline}
Further details can be found in Martens and Grosse (2015).\cite{martens2015optimizing}

The original KFAC derivation assumed the density to be estimated was normalized, but we wish to extend it to stochastic reconfiguration for unnormalized wavefunctions. In Appendix~\ref{sec:ngd_and_sr}, we show that if we only have access to an unnormalized wavefunction, terms in the FIM can be expressed as:
\[
\mathcal{F}_{ij} \propto \mathbb{E}_{p(\overallState)}[(\mathcal{O}_i-\mathbb{E}_{p(\overallState)}[\mathcal{O}_i])(\mathcal{O}_j-\mathbb{E}_{p(\overallState)}[\mathcal{O}_j])]
\]
where $\mathcal{O}_i = \frac{\partial \mathrm{log}|\psi|}{\partial x_i}$. The terms in the FIM for the weights of a linear neural network layer would then be:
\begin{widetext}
\begin{align*}
	\expectp{\frac{\partial \mathrm{log}p(\overallState)}{\partial \mathrm{vec}(\mathbf{W}_\ell)}  \frac{\partial \mathrm{log}p(\overallState)}{\partial \mathrm{vec}(\mathbf{W}_\ell)}^T} &\propto \expectp{\left(\mathbf{a}_\ell\otimes\mathbf{e}_\ell-\mathbb{E}_{p(\overallState)}[\mathbf{a}_\ell\otimes\mathbf{e}_\ell]\right)\left(\mathbf{a}_\ell\otimes\mathbf{e}_\ell-\mathbb{E}_{p(\overallState)}[\mathbf{a}_\ell\otimes\mathbf{e}_\ell]\right)^T} \\
	&= \expectp{\left(\mathbf{a}_\ell\otimes\mathbf{e}_\ell\right)\left(\mathbf{a}_\ell\otimes\mathbf{e}_\ell\right)^T} - \expectp{\mathbf{a}_\ell\otimes\mathbf{e}_\ell} \expectp{\mathbf{a}_\ell\otimes\mathbf{e}_\ell}^T \\
\end{align*}
\end{widetext}

\begin{figure}
    \centering
    \includegraphics[width=\columnwidth]{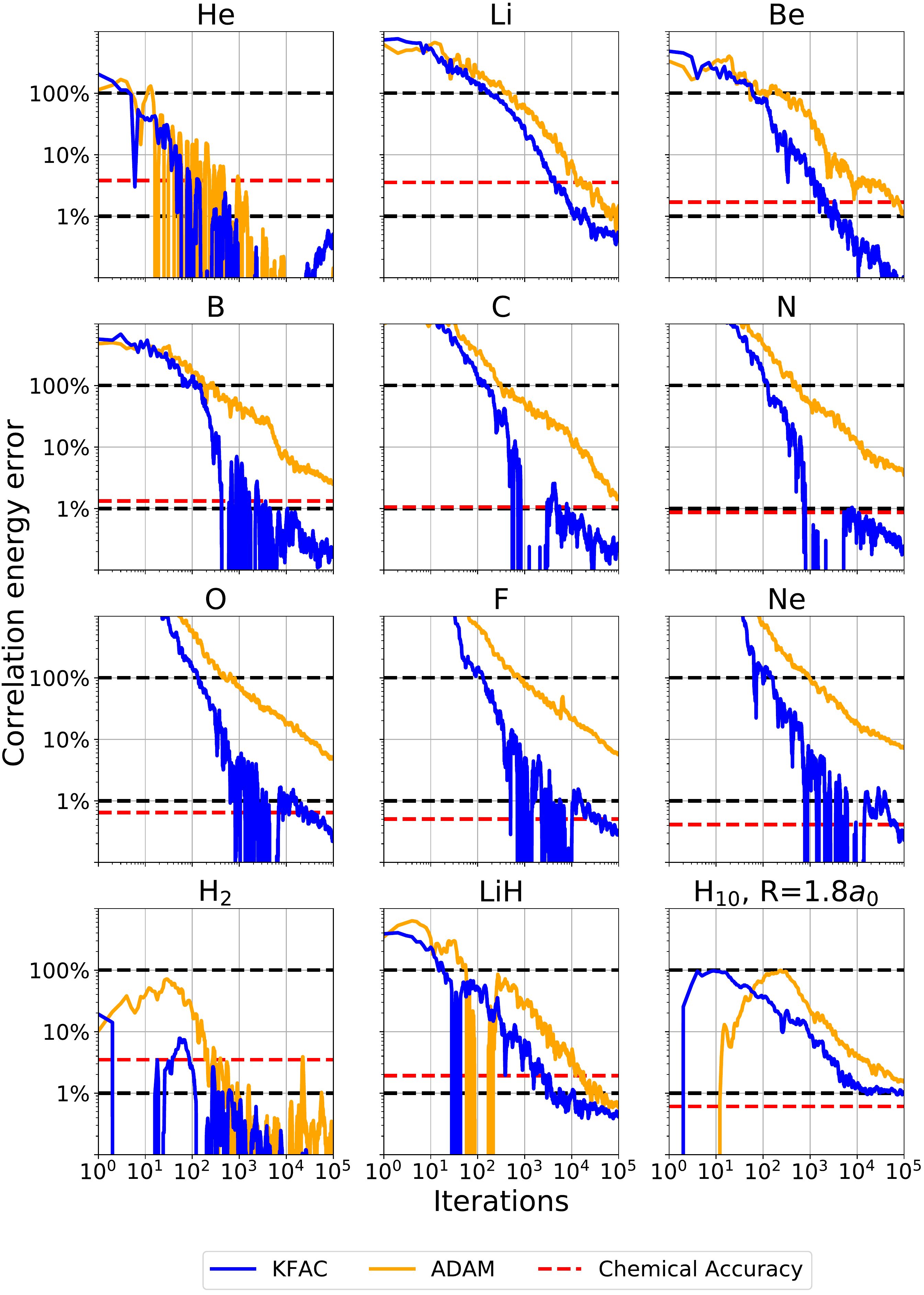}
    \caption{optimization progress for first-row atoms, H$_2$, LiH and the hydrogen chain with KFAC (blue) vs. ADAM (orange). The qualitative advantage of KFAC is clear. For clarity, the median energy over the last 10\% of iterations is shown. Note that the small overshoot with KFAC between $10^3$ and $10^4$ iterations is due to the slow equilibration of the MCMC chain and goes away with a larger Metropolis-Hastings proposal step size.}
    \label{fig:kfac_adam}
\end{figure}

We use a similar approximation for the inverse to that of conventional KFAC, replacing the uncentered second moments with mean-centered covariances:
\begin{multline}
	\expectp{\frac{\partial \mathrm{log}p(\overallState)}{\partial \mathrm{vec}(\mathbf{W}_\ell)}  \frac{\partial \mathrm{log}p(\overallState)}{\partial \mathrm{vec}(\mathbf{W}_\ell)}^T} \approx \\
	\expectp{\mathbf{\hat{a}}_\ell\mathbf{\hat{a}}_\ell^T}^{-1}\otimes\expectp{\mathbf{\hat{e}}_\ell\mathbf{\hat{e}}_\ell^T}^{-1},
\end{multline}
where
\begin{align*}
	\mathbf{\hat{a}}_\ell &= \mathbf{a}_\ell - \expectp{\mathbf{a}_\ell}, \\
	\mathbf{\hat{e}}_\ell &= \mathbf{e}_\ell - \expectp{\mathbf{e}_\ell}.
\end{align*}
We illustrate the advantage of using KFAC over more commonly used stochastic first order optimization methods for neural networks in Fig.~\ref{fig:kfac_adam}.

\section{Results}

Here we evaluate the performance of the FermiNet on a variety of problems in chemistry and electronic structure. Further details on the exact architectures and training procedures for the FermiNet and baselines can be found in Appendix~\ref{sec:experimental_setup}.

\subsection{Slater-Jastrow versus FermiNet Ans{\"a}tze}

To demonstrate the expressive power of the FermiNet, we first investigated its performance relative to the more conventional Slater-Jastrow and Slater-Jastrow-backflow Ans{\"a}tze with varying numbers of determinants:
\begin{align}
    \Psi_{\textrm{SJ}} &= e^{J(\{\mathrm{r}_i\})} \sum_k \omega_k 
                            \det\left[\phi^{k\uparrow}_i(\mathbf{r}^{\uparrow}_j) \right]
                            \det\left[\phi^{k\downarrow}_i(\mathbf{r}^{\downarrow}_j) \right] \\
    \Psi_{\textrm{SJB}} &= e^{J(\{\mathrm{r}_i\})} \sum_k \omega_k 
                            \det\left[\phi^{k\uparrow}_i(\mathbf{q}^{\uparrow}_j) \right]
                            \det\left[\phi^{k\downarrow}_i(\mathbf{q}^{\downarrow}_j) \right]
\end{align}
where $\{{\phi^{k\alpha}_i(\mathbf{r}_j})\}$ is a set of single-particle orbitals, typically obtained from a Hartree--Fock or density functional theory calculation, and the Jastrow factor, $J$ is a function of the electron and nuclear coordinates. The Slater-Jastrow-backflow wavefunction Ansatz replaces the electron coordinates in the orbitals with a set of collective coordinates, given by $\mathbf{q}_i = \mathbf{r}_i + \xi_i(\{\mathbf{r}_j\})$, where the backflow functions $\xi_i$ depend on electron and nuclear coordinates and contain additional optimizable parameters.

In addition to conventional Slater-Jastrow and Slater-Jastrow backflow wavefunctions, we also compare against neural network versions. Rather than using Hartree-Fock orbitals, a closed-form Jastrow factor, and a backflow transform with only a few free parameters, our Slater-Jastrow-backflow network uses residual neural networks to represent the one-electron orbitals, Jastrow factor and backflow transform, making it much more flexible. The determinant part of the Slater-Jastrow network amounts to removing the two-electron stream and interactions between the one-electron streams from FermiNet. We used the conventional backflow transformation (Equation~\ref{eqn:backflow}), in which the orbitals depend on a single three-dimensional linear combination of electron position vectors and a nonlinear function of interparticle distances. Further details are provided in Appendix~\ref{sec:SJ_nets}.

To fairly compare our calculations against previous work, we first looked at single-determinant Ans\"atze for first-row atoms. Figure~\ref{fig:compare_det}a compares the FermiNet results with numbers already available in the literature.\cite{seth2011quantum} The neural network Slater-Jastrow Ansatz already outperforms the numbers from the literature by a few milli-Hartrees (mE$_h$), which could be due to the lack of basis set approximation error when using a neural network to represent the orbitals and a flexible Jastrow factor. 
The FermiNet cuts the error relative to the Slater-Jastrow Ansatz without backflow by almost an order of magnitude, and more than a factor of two relative to the Slater-Jastrow-Backflow Ansatz. Just a single FermiNet determinant is sufficient to come within a few mE$_h$ of chemical accuracy, defined as 1 kcal/mol (1.594 mE$_h$), which is the typical standard for a quantum chemical calculation to be considered ``correct."

Not only is the FermiNet a significant improvement over the Slater-Jastrow Ansatz with one determinant, but only a few FermiNet determinants are necessary to saturate performance. Figure~\ref{fig:compare_det}b shows the Slater-Jastrow network and FermiNet energies of the nitrogen and carbon monoxide molecules as functions of the number of determinants. As FCI calculations are impractical for these systems, we compare against the unrestricted coupled cluster singles, doubles, and perturbative triples method (CCSD(T)) in the complete basis set (CBS) limit to provide a comparable baseline for both systems. As the Slater-Jastrow network optimizes all orbitals separately, the results from the Slater-Jastrow network should be a lower bound on the performance of a Slater-Jastrow Ansatz with a given number of determinants. As expected, the Slater-Jastrow network is still far from the accuracy of CCSD(T) at 64 determinants. The 64-determinant FermiNet, in contrast, comes within a few mE$_h$ of CCSD(T). While the Slater-Jastrow-backflow Ansatz with large numbers of determinants did not completely converge, the trend is clear that the FermiNet cuts the error roughly in half. The FermiNet energies begin to plateau after only a few tens of determinants, suggesting that large linear combinations of FermiNet determinants are not required. Despite recent advances in optimal determinant selection,\cite{giner2016qmc,dash2018perturbatively} conventional Slater-Jastrow VMC calculations typically require tens of thousands of determinants for systems of this size and rarely match CCSD(T) accuracy even then.

\subsection{Equilibrium Geometries}

Tables \ref{tab:energy} and \ref{tab:energy_molecules} show that the same 16-determinant FermiNet with the same training hyperparameters generalizes well to a wide variety of atoms and diatomic and small organic molecules, while Figure~\ref{fig:kfac_adam} shows the optimization progress over time for many of these systems. As a baseline, we used a combination of experimental and exact computational results where available,\cite{chakravorty1993ground, cencek2000benchmark, filippi1996multiconfiguration} and all-electron CBS CCSD(T) otherwise. On all atoms, as well as LiH, Li$_2$, methane and ammonia, the FermiNet error was within chemical accuracy. In comparison, energies from VMC using a conventional Slater-Jastrow-backflow Ansatz for first-row atoms\cite{seth2011quantum} are uniformly worse than the FermiNet, despite using at least an order of magnitude more determinants. The VMC-based FermiNet energies are more comparable in quality to diffusion Monte Carlo (DMC), which is typically much more accurate than VMC. On molecules as large as ethene (C$_2$H$_4$) we recover over 99\% of the correlation energy, while for larger systems like methylamine (CH$_3$NH$_2$), ozone (O$_3$), ethanol (C$_2$H$_5$OH) and bicyclobutane (C$_4$H$_6$) the percentage of correlation energy recovered declines gradually to $\sim$97\% -- still remarkably good for a variational calculation. Bicyclobutane is an especially challenging system due to its high ring strain and large number of electrons. 

\begin{figure}
	\centering
	\includegraphics[width=\columnwidth]{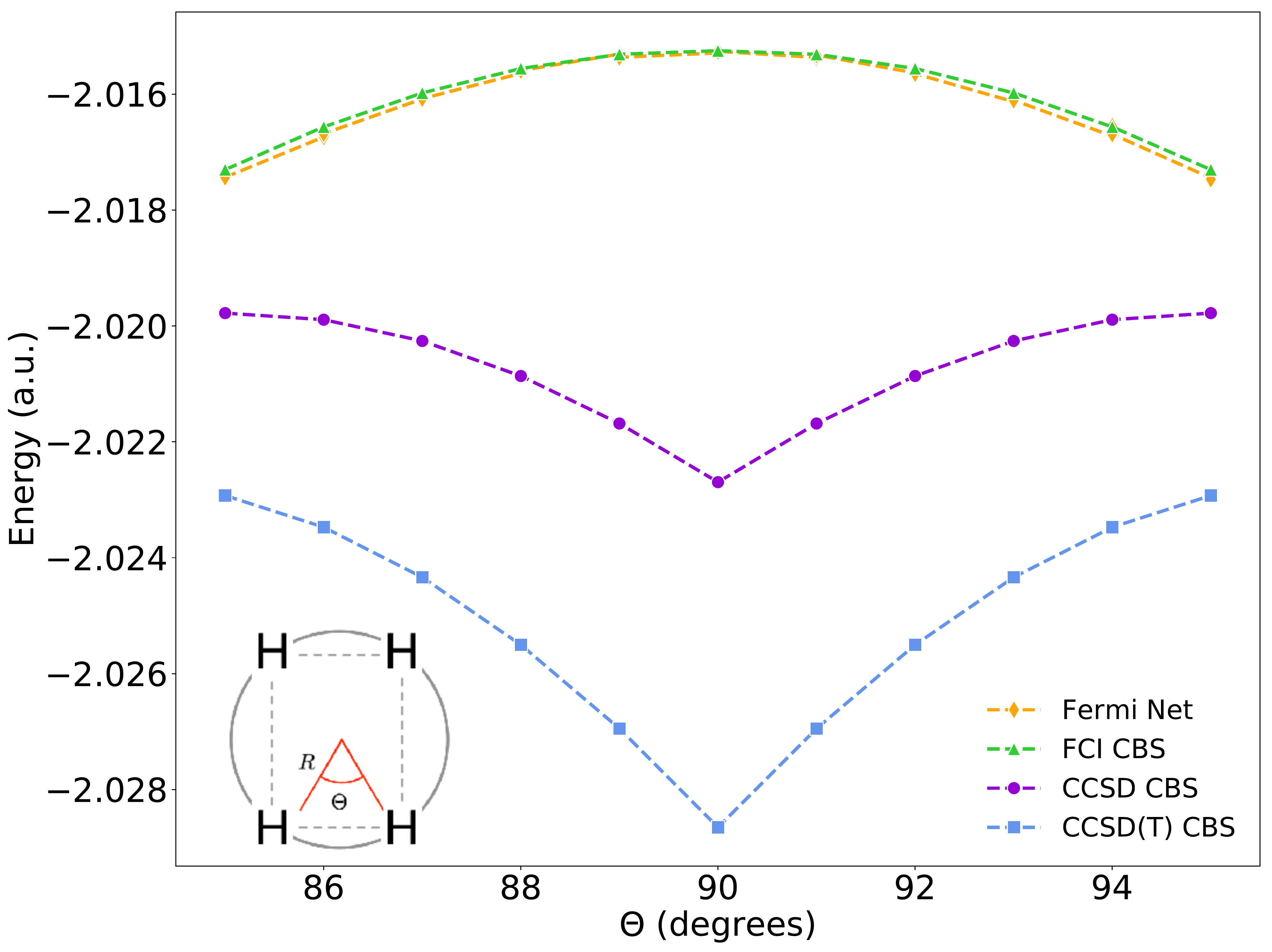}
	\vspace{0.5cm}
	\caption{
	The H$_4$ rectangle, $R=3.2843a_0$. Coupled cluster methods incorrectly predict a cusp and energy minimum at $\Theta=90^\circ$, while the FermiNet approach agrees with exact FCI results.
	}\label{fig:h4}
\end{figure}

We also compare against CCSD(T) in a finite basis set in Table~\ref{tab:energy_molecules}, and find that in all cases the FermiNet is {\em more} accurate than CCSD(T) in the largest basis set we could practically run calculations on (quintuple $\zeta$ for most systems, quadruple $\zeta$ for large systems). This suggests that a comparable extrapolation of FermiNet results could match or even exceed the accuracy of CCSD(T). As the FermiNet works directly in the continuum and does not depend on a basis set, the natural equivalent would be extrapolation to the limit of infinitely-wide layers in the one-electron stream. Our analysis of the FermiNet with different numbers of layers and layer widths in Sec.~\ref{sec:feature_ablation} shows that the error appears to decrease polynomially with layer width.

We also computed the first ionization potentials, $E(X^{+})-E(X)$ for element $X$, and electron affinities, $E(X)-E(X^{-})$, for first-row atoms (Table~\ref{tab:energy}) and compare to experimental data\cite{klopper2010submev} with relativistic effects removed. Agreement with experiment is excellent (mean absolute error of 0.09 mE$_h$ for ionization potentials and 0.66 mE$_h$ for electron affinities), demonstrating that the FermiNet Ansatz is capable of representing charged and neutral species with similar accuracy.

There are many possible causes for the decline in the percent of correlation energy recovered for large systems like bicyclobutane. It may be that the FermiNet has issues with size-extensivity for larger systems. On the other hand, the FermiNet outperforms CCSD(T) in a fixed basis set, and the exponential Ansatz used by coupled cluster is size extensive, suggesting that the issue may instead be the finite width of our neural network layers. In fact, our results are with a fixed-width network, while the total number of basis functions grows with the system size for coupled cluster, meaning the coupled cluster Ansatz becomes more expressive for larger systems while the FermiNet stays fixed. Other avenues for improvement include increasing the batch size/number of walkers, improving the MCMC chain mixing and optimization efficiency, and increasing the number of determinants.

\subsection{The H$_4$ Rectangle}

\begin{figure}
	\centering
	\includegraphics[width=\columnwidth]{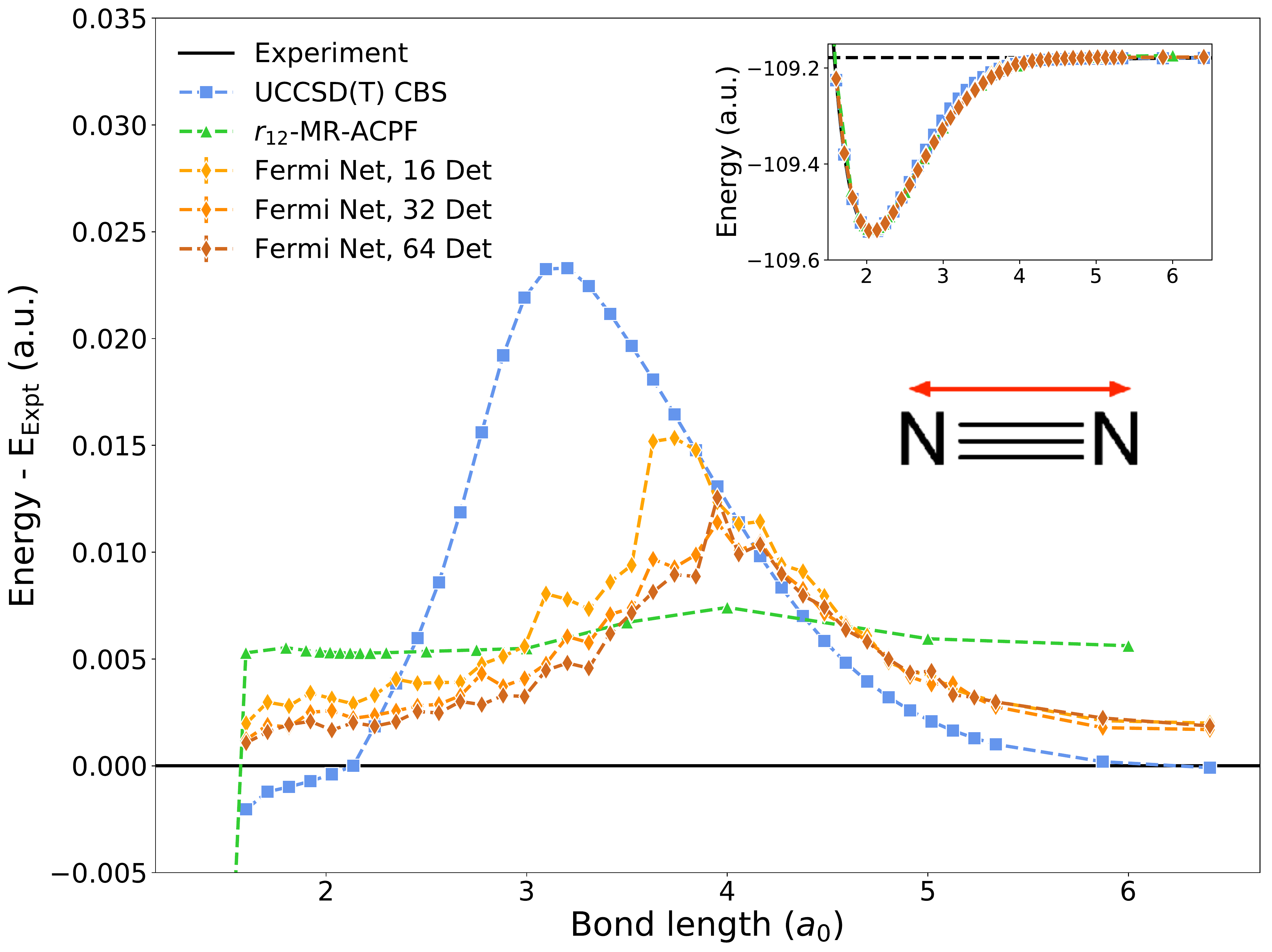}
	\vspace{0.5cm}
	\caption{
	The dissociation curve for the nitrogen triple-bond. The difference from experimental data\cite{leroy2006accurate} is given in the main panel. In the region of largest UCCSD(T) error, the FermiNet prediction is comparable to highly-accurate $r_{12}$-MR-ACPF results.\cite{gdanitz1998accurately} 
	}\label{fig:n2}
\end{figure}

\begin{figure}
	\centering
	\includegraphics[width=\columnwidth]{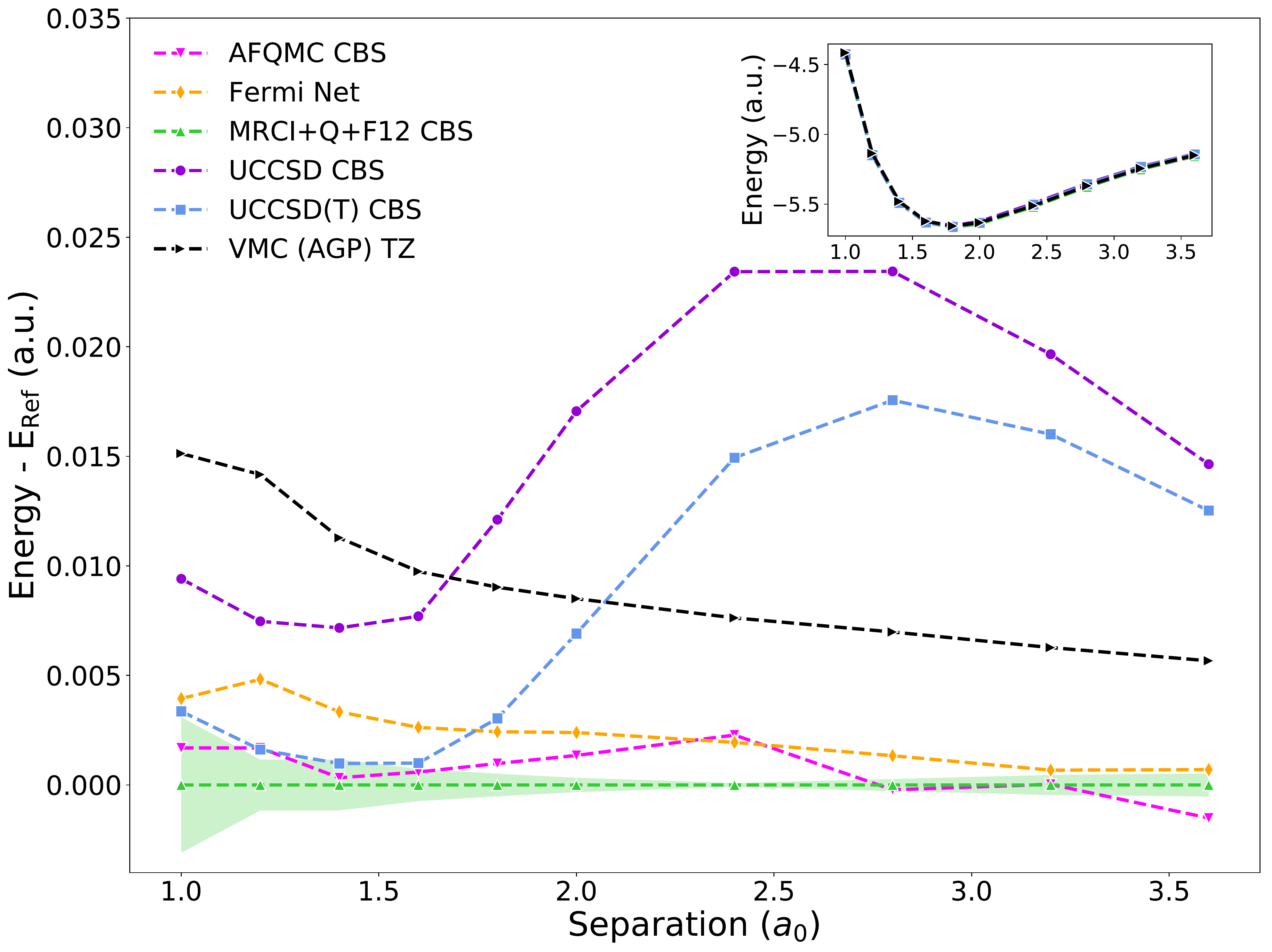}
	\vspace{0.5cm}
	\caption{
	The H$_{10}$ chain. All energies except the FermiNet are taken from Motta et al.~(2017)\cite{motta2017towards}. The absolute energies (inset) cannot be distinguished by eye. The difference from highly accurate MRCI+Q+F12 results are shown in the main panel, where the shaded region indicates an estimate of the basis-set extrapolation error. The errors in the coupled cluster and conventional VMC energies are large at medium atomic separations but the FermiNet remains comparable to AFQMC at all separations. See also Appendix~\ref{sec:hn_consistency} for data on larger separations.}
	\label{fig:h10}
\end{figure}

While CCSD(T) is exceptionally accurate for equilibrium geometries, it often fails for molecules with low-lying excited states or stretched, twisted or otherwise out-of-equilibrium geometries. Understanding these systems is critical for predicting many chemical properties. A model system small enough to be solved exactly by FCI for which coupled cluster fails is the rectangle of four hydrogen atoms, parametrized by the distance $R$ of the atoms from the centre and the angle $\theta$ between neighbouring atoms.\cite{vanvoorhis2000benchmark} FCI shows that the energy varies smoothly with $\theta$ and is maximized when the atoms are at the corners of a square ($\theta = 90^{\text{o}}$). The coupled cluster results are non-variational, predicting energies too low  by several milli-hartree, and qualitatively incorrect, predicting an energy {\em minimum} with a non-analytic downward-facing cusp at $90^{\text{o}}$, caused by a crossing of two Hartree--Fock states with different symmetries.\cite{burton2016holomorphic} Figure~\ref{fig:h4} shows that the FermiNet does not suffer from the same errors as coupled cluster and is in essentially perfect agreement with FCI. We attribute the small discrepancy between the FermiNet and FCI energies to errors arising from the basis set extrapolation used for the FCI energies.

\subsection{The Nitrogen Molecule}
\label{sec:nitrogen_dimer}

A problem more relevant to real chemistry that troubles coupled cluster methods is the dissociation of the nitrogen molecule. The triple bond is challenging to describe accurately and the stretched N$_2$ molecule has several low-lying excited states, leading to errors when using single-reference coupled cluster methods.\cite{lyakh2011multireference} Experimental values for the dissociation potential have been reconstructed from spectroscopic measurements using the Morse/Long-range potential.\cite{leroy2006accurate} These closely match calculations using the $r_{12}$-MR-ACPF method,\cite{gdanitz1998accurately} which is highly accurate but scales factorially. A comparison between unrestricted CCSD(T), the FermiNet, and these high-accuracy methods is given in Figure~\ref{fig:n2}. The total FermiNet error is significantly smaller than UCCSD(T), and in the region of largest UCCSD(T) error the FermiNet reaches accuracy comparable to $r_{12}$-MR-ACPF, but scales much more favourably with system size. Increasing the number of determinants in the FermiNet improves performance up to a point, but not beyond 32 determinants, again suggesting that the bottleneck to performance is not size-consistency. While coupled cluster could in theory be made more accurate by extending to full triples or quadruples, or using multireference methods, CCSD(T) is generally considered the largest coupled cluster approximation that can reasonably scale beyond small molecules. This shows that, without any specific tuning to the system of interest, the FermiNet is a clear improvement over single-reference coupled cluster for modelling a strongly correlated real-world chemical system.

\begin{figure}
    \centering
    \includegraphics[width=\columnwidth]{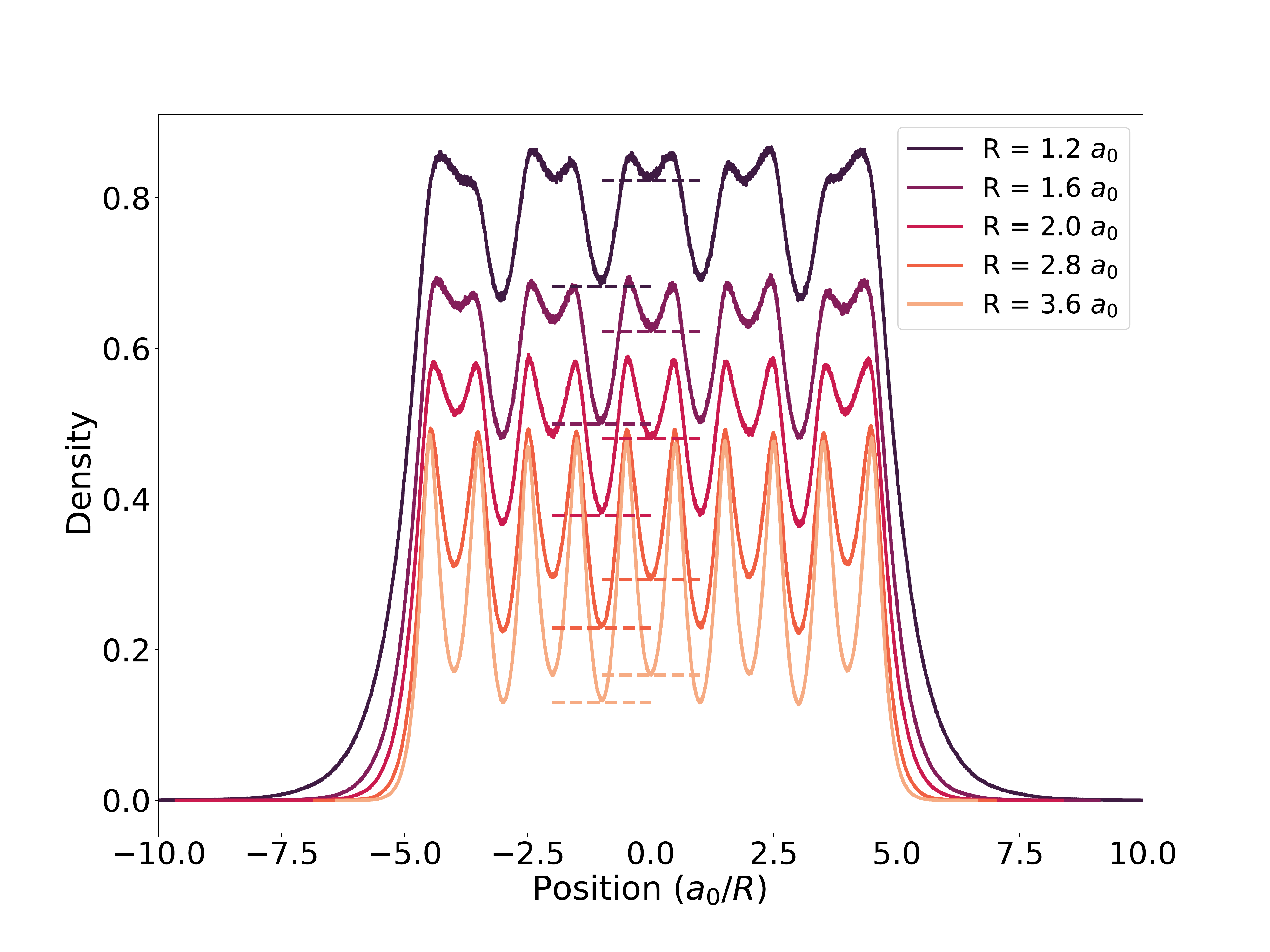}\\
    \caption{Electron dimerization in the hydrogen chain. The gap between alternating minima of the density shrinks with increasing nuclear separation.}
    \label{fig:dimerization}
\end{figure}

\subsection{The Hydrogen Chain}
\label{sec:hydrogen_chain}

Finally, we investigated the performance of the FermiNet on the evenly-spaced linear hydrogen chain. The hydrogen chain is of great interest as a system that bridges model Hamiltonians and real material systems and may undergo an insulator-to-metal transition as the separation of the atoms is decreased. Consequently, results obtained using a wide range of many-electron methods have been rigorously evaluated and compared.\cite{motta2017towards} We compare the performance of the FermiNet against many of these methods in Figure~\ref{fig:h10}. Of the two projector QMC methods studied by Motta \emph{et al}.~, AFQMC gave slightly better results than lattice regularized DMC and so we omit the latter for clarity. Without changing the network architecture or hyperparameters, we are again able to outperform coupled cluster methods and conventional VMC and obtain results competitive with state-of-the-art approaches.

\section{Analysis}

Here we provide an analysis of the performance of the FermiNet, looking at scaling with system size, network size, and visualising quantities beyond just total energy of the system.

\subsection{Electron Densities}

\begin{figure}[t]
    \centering
    \includegraphics[width=\columnwidth]{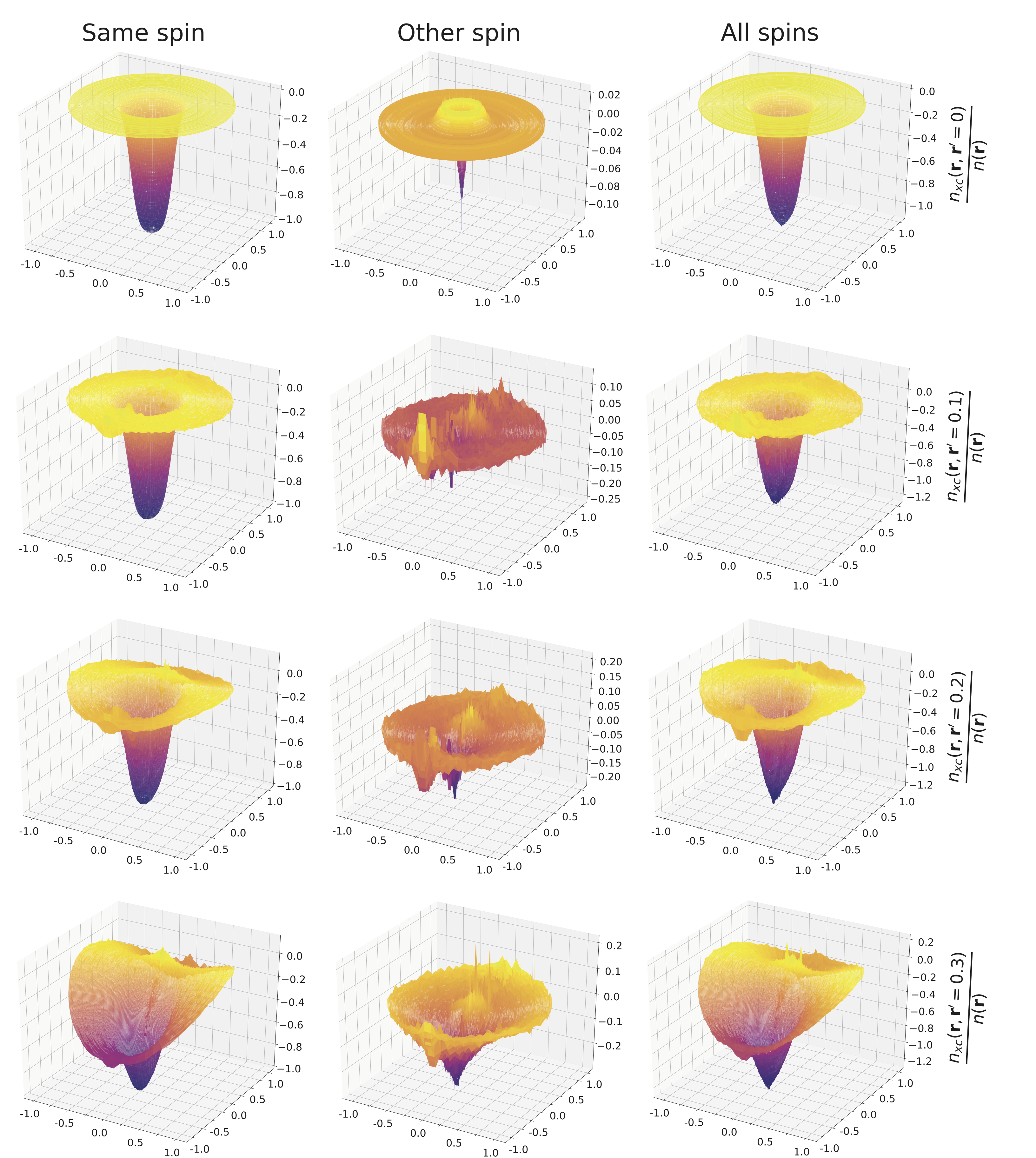}\\
    \caption{The pair-correlation function $1 - g(\mathbf{r}, \mathbf{r}')$ for the neon atom, where $n(\mathbf{r}, \mathbf{r}') = n(\mathbf{r})g(\mathbf{r}, \mathbf{r}')n(\mathbf{r}')$. Different columns show the hole for electrons of the same spin (left), different spins (middle), or all electrons (right). Different rows show the hole when $\mathbf{r}'$ is between 0 and 0.3 Bohr radii from the nucleus.}
    \label{fig:xc_hole}
\end{figure}

One advantage of VMC over other ab-initio electronic structure methods is the ease of evaluating expectation values of arbitrary observables. For instance, forces are significantly easier to calculate with VMC than projector QMC methods.\cite{badinski2010methods} To illustrate the quality of the FermiNet Ansatz for observables other than energy, we analyzed the one- and two-electron densities. The electron-electron and electron-nuclear cusps for the helium atom are investigated in Appendix~\ref{sec:cusps}.

For the hydrogen chain, we computed the one-electron density $n(\mathbf{r})$ at different nuclear separations, shown in Figure~\ref{fig:dimerization}. Consistent with many other electronic structure methods,\cite{motta2019ground}  we found that the electron density undergoes a dimerization --- the density clusters around pairs of nuclei --- and the effect becomes stronger with less separation between nuclei. Dimerization is a hallmark of electronic structure in insulators, and understanding when and where it occurs helps understand metal-insulator phase transitions in materials.

Additionally, we investigated the two-electron density $n(\mathbf{r}, \mathbf{r}')$ for the neon atom (Figure~\ref{fig:xc_hole}). Understanding the behaviour of the two-electron density is important for many electronic structure methods, for instance for analysing functionals for DFT.\cite{hood1997quantum} What is interesting about the two-electron density is how it {\em differs} from the product of one-electron densities, $n(\mathbf{r})n(\mathbf{r}')$. This can be expressed in terms of the exchange-correlation hole, $n_{xc}(\mathbf{r}, \mathbf{r}')$, defined such that $n(\mathbf{r},\mathbf{r}') = [n(\mathbf{r}) + n_{xc}(\mathbf{r},\mathbf{r}')]n(\mathbf{r}')$, or in terms of the pair-correlation function, $g(\mathbf{r}, \mathbf{r}')$, defined by $n(\mathbf{r},\mathbf{r}') = n(\mathbf{r}) g(\mathbf{r},\mathbf{r}')n(\mathbf{r}')$. As most of the density is concentrated near $\mathbf{r}=\mathbf{0}$, $n_{xc}(\mathbf{r},\mathbf{r'})$ is very strongly peaked near $\mathbf{r}=\mathbf{0}$, obscuring its other features. We therefore show $\frac{n_{xc}(\mathbf{r}, \mathbf{r}')}{n(\mathbf{r})} = 1 - g(\mathbf{r}, \mathbf{r}')$ in Figure~\ref{fig:xc_hole}. This behaves as expected when $\mathbf{r}$ is close to $\mathbf{r}'$, showing that, at least for first and second order statistics, the FermiNet Ansatz is smooth and well-behaved.

\subsection{Scaling and Computation Time}

\begin{figure}
    \centering
    \includegraphics[width=\columnwidth]{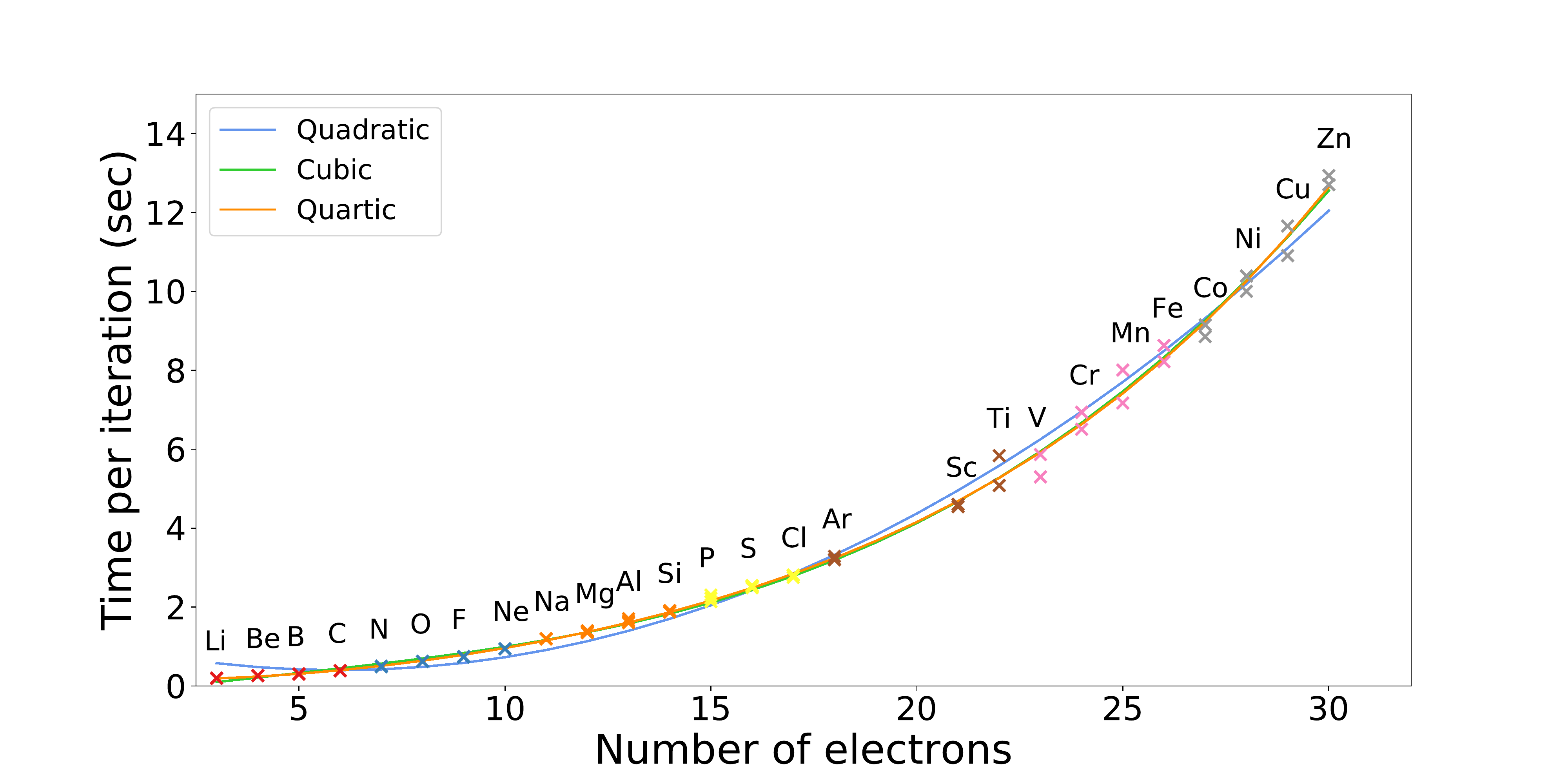}
    \caption{Comparison of the runtime for one optimization iteration on atoms up to zinc. Polynomial regressions up to fourth order are fit to the data. The small difference between the cubic and quartic fit suggests that the determinant computation is not the dominant factor at this scale.}
    \label{fig:scaling}
\end{figure}

\begin{figure*}
    \centering
    \includegraphics[width=\textwidth]{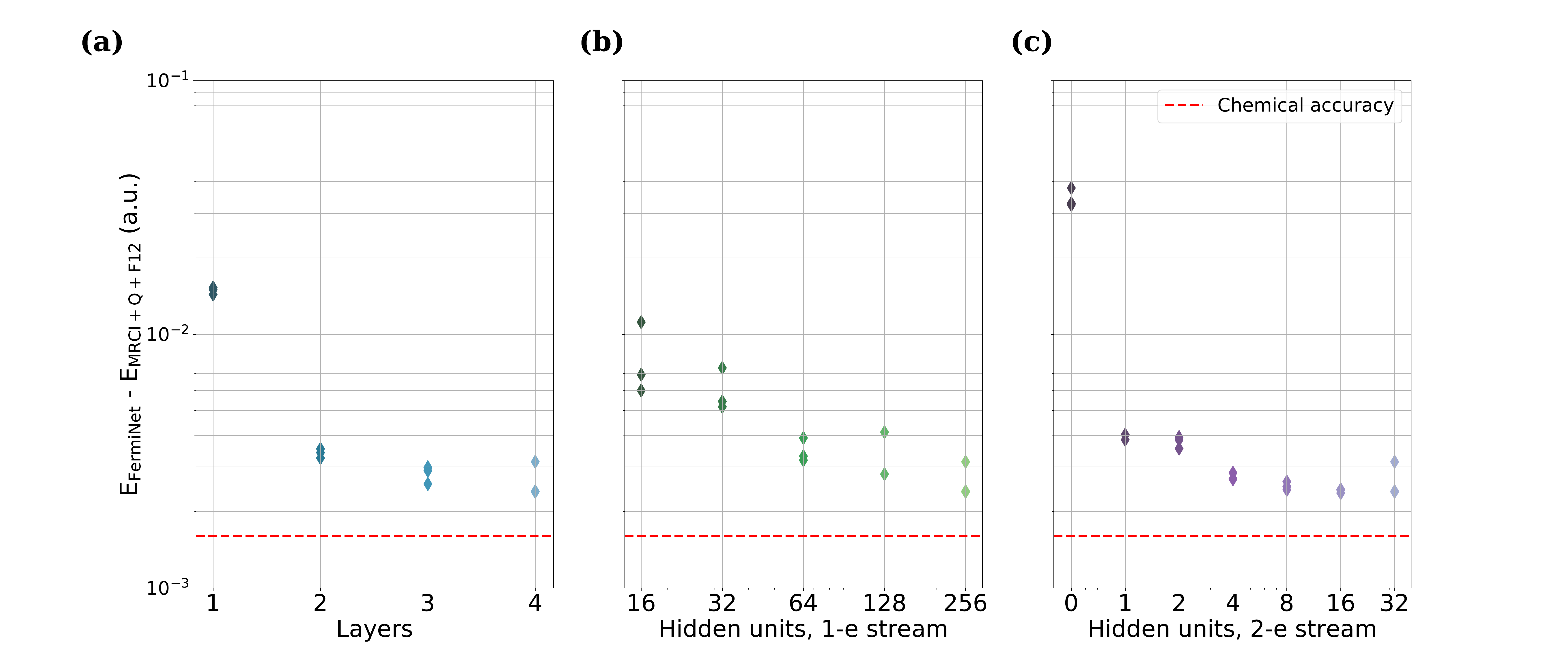}\\
    \caption{Effects of network architecture on FermiNet performance on the hydrogen chain H$_{10}$ with separation $R=2.0 a_0$. Each point is one run of the same model. \textbf{(a)}: Effect of network depth. The marginal improvement with 4 layers is small but not zero. \textbf{(b)}: Effect of number of hidden units in the one-electron stream. There is a continuous improvement with wider layers, with the error decreasing roughly as $\mathcal{O}(N^{-0.395\pm 0.067})$. \textbf{(c)}: Effect of number of hidden units in the two-electron stream. The accuracy plateaus above 16 units.}
    \label{fig:h10_ablation}
\end{figure*}

One of our main claims about the FermiNet is that it scales favorably compared to other ab-initio quantum chemistry methods. The ability to run at all on systems the size of bicyclobutane proves the FermiNet scales more favorably than exact methods like FCI, but the scaling relative to other approximate methods is a more subtle question. Both the size of the FermiNet (number of hidden units, number of layers, number of determinants) and the number of training iterations required to reach a certain level of accuracy are unknown, and likely depend on the system being studied. What can be quantified is the computational complexity of a single iteration of training, which can be seen as a lower bound on the computational complexity of training the FermiNet to a certain level of accuracy.

For a system with $N_e$ electrons, $N_a$ atoms and a FermiNet with $L$ hidden layers, $n_1$ ($n_2)$ hidden units per one-electron (two-electron) layer and $n_k$ determinants, evaluating the one-electron stream of the network scales as $\mathcal{O}(N_e(N_a + L (n_1^2 + n_1n_2)))$, evaluating the two-electron stream scales as $\mathcal{O}(N_e^2 L n_2^2)$,  evaluating the orbitals and envelope scales as $\mathcal{O}(N_e^2 n_k (n_1 + N_a))$, and evaluating the determinants scales as $\mathcal{O}(N_e^3 n_k)$, so the determinant calculation should dominate as $N_e$ grows for a fixed network architecture determined by $\{L, n_1, n_2, n_k\}$. While evaluating the gradient of a function has the same asymptotic complexity as evaluating the function, evaluating the local energy scales with an additional multiplicative factor of $N_e$, as computing the Laplacian has the same complexity as computing the Hessian with respect to the inputs, giving an asymptotic complexity of $\mathcal{O}(N_e^4 n_k)$ as $N_e$ grows. A Markov Chain Monte Carlo (MCMC) step for sampling from $\psi^2$ also has the same asymptotic complexity as network evaluation for all-electron moves, or similar complexity to Laplacian calculation for single-electron moves if all electrons are moved in each loop of training.

The number of total parameters scales as $\mathcal{O}(N_a n_1 + L(n_1^2 + n_1 n_2 + n_2^2) + N_e n_k (n_1 + N_a) + n_k)$ (see Table~\ref{tab:parameters} for the exact shapes for each parameter). Note that, other than the orbital shaping and envelope parameters, there is no direct dependence on $N_e$. KFAC requires a matrix inversion for each Kronecker-factorized block of the approximate FIM, which scales as $\mathcal{O}(m^3+n^3)$ for a linear layer with $m$ inputs and $n$ outputs. For the FermiNet, this works out to a scaling of $\mathcal{O}(N_a^3 n_1^3 + L(n_1^3 + n_2^3) + (N_e n_k n_1)^3 + (N_e n_k N_a)^3 + n_k^3)$. Combining the MCMC steps, local energy calculation and KFAC update together gives an overall quartic asymptotic scaling with system size for a single step of wavefunction optimization. We emphasize that the analysis here treats system size, network size and number of samping steps independently, and that the exact dependence of network size and sampling parameters on system size to achieve constant accuracy requires further investigation.

We give an empirical analysis of the scaling of iteration time in Figure~\ref{fig:scaling} on atoms from lithium to zinc, using the default training configuration with 8 GPUs. For larger atoms, we were not able to run optimization to convergence, but we were able to execute enough updates to get an accurate estimate of the timing for a single iteration consisting of 10 MCMC steps, a local energy and gradient evaluation and a KFAC update. Fitting polynomials of different order to the curve, we find a cubic fit is able to accurately match the scaling, suggesting that for systems of this size the computation is dominated by the $O(N^2)$ evaluation of the two-electron stream of the FermiNet, while the determinant only becomes dominant for much larger systems.

\subsection{Feature Ablation and Network Architectures}
\label{sec:feature_ablation}

\begin{table}[!t]
    \centering
    \begin{tabular}{c|cc}
         $\Delta$E (mE$_h$) & without $\mathbf{r}_{ij}$ & with $\mathbf{r}_{ij}$ \\
         \hline
         without $|\mathbf{r}_{ij}|$ & 89.7 & 28.4 \\
         with $|\mathbf{r}_{ij}|$ & 1.2 & \textbf{0.8}
    \end{tabular}
    \caption{\textbf{Performance of the FermiNet on the oxygen atom with input features removed.} All configurations without the electron-nuclear distances $|\mathbf{r}_{iI}|$ were numerically unstable and diverged. All numbers are relative to Chakravorty (1993).\cite{chakravorty1993ground}}
    \label{tab:input_ablation}
\end{table}

There are many free parameters in the FermiNet architecture that must be chosen to maximize accuracy for a given amount of computation. To illustrate the effect of different architectural choices, we removed many features, layers and hidden units from the FermiNet and investigated how the performance decayed. The FermiNet has 4 distinct input features: the nuclear coordinates $\mathbf{r}_{iI} = \mathbf{r}_i - \mathbf{R}_I$ and nuclear distances $|\mathbf{r}_{iI}|$, which are inputs to the one-electron stream, and the interelectron coordinates $\mathbf{r}_{ij} = \mathbf{r}_i - \mathbf{r}_j$ and interelectron distances $|\mathbf{r}_{ij}|$, which are inputs to the two-electron stream. We compared the accuracy of the FermiNet with and without these features on the oxygen atom in Table~\ref{tab:input_ablation}. All networks included the nuclear coordinates. Without the nuclear distances, the network became unstable and training crashed, possibly due to the inability to accurately capture the electron-nuclear cusp conditions. When including interelectron features, most of the increase in accuracy was due to the distances $|\mathbf{r}_{ij}|$, while the coordinates $\mathbf{r}_{ij}$ also improved accuracy, though not by as large an amount. This shows that all input features contributed towards stability and accuracy, especially the distance features. Even though a smooth neural network can approximate the non-smooth cusps to high precision (although not perfectly), by including distances, which are non-smooth at zero, we can make the wavefunction significantly easier to approximate.

To understand the effect of the size and shape of the network, we compared the FermiNet with different numbers of layers and hidden units on the hydrogen chain H$_{10}$. The results are presented in Figure~\ref{fig:h10_ablation}. When increasing the number of layers, the overall accuracy increases as more layers are added, but the difference from 3 to 4 layers is only on the order of 1 mE$_h$, suggesting that the gains from additional layers would be minor. When adding more hidden units to the one-electron stream but keeping 32 units in the two-electron stream, the accuracy increases uniformly with more units. Based on a linear regression of the log-errors relative to MRCI+Q+F12, and using bootstrapping to generate error bars, the error scales with the number of hidden units in the one-electron stream as $\mathcal{O}(N^{-0.395\pm 0.067})$. This means we would expect around 760 hidden units to be needed to reach chemical accuracy on the hydrogen chain. For the two-electron stream, the improvement with more units quickly saturates. In fact, going from 16 to 32 hidden units seems to make the results slightly noisier. This suggests that increasing the width of the one-electron stream, more than increasing the width of the two-electron stream or the total depth, is the most promising route to increasing overall accuracy of the FermiNet.

\section{Discussion}

We have shown that antisymmetric neural networks can be constructed and optimized to enable high-accuracy quantum chemistry calculations of challenging systems. The Fermionic Neural Network makes the simple and straightforward VMC method competitive with DMC, AFQMC and CCSD(T) methods for equilibrium geometries and better than CCSD(T) for many out-of-equilibrium geometries. Importantly, one network architecture with one set of training parameters has been able to attain high accuracy on every system examined. The use of neural networks means that we do not have to choose a basis set or worry about basis-set extrapolation, a common source of error in computational quantum chemistry. There are many possible applications of the FermiNet beyond VMC, for instance as a trial wavefunction for projector QMC methods. We expect further work investigating the tradeoffs of different antisymmetric neural networks and optimization algorithms to lead to greater computational efficiency, higher representational capacity, and improved accuracy on larger systems. This has the potential to bring to quantum chemistry the same rapid progress that deep learning has enabled in numerous fields of artificial intelligence.

\begin{acknowledgments}
We would like to thank J. Jumper, J. Kirkpatrick, M. Hutter, T. Green, N. Blunt, S. Mohamed and A. Cohen for helpful discussions, B. McMorrow for providing data, J. Martens and P. Buchlovsky for assistance with code, and A. Obika, S. Nelson, C. Meyer, T. Back, S. Petersen, P. Kohli, K. Kavukcuoglu and D. Hassabis for support and guidance. Additional thanks to the rest of the DeepMind team for support, ideas and encouragement.

D.P.~and J.S.S.~ contributed equally to this work.
Correspondence and requests for materials should be addressed to D.P.~(pfau@google.com).
\end{acknowledgments}

\appendix

\section{Experimental Setup}
\label{sec:experimental_setup}

\begin{table*}
    \begin{tabular}{ccccl}\hline\hline
        symbol                                      & dimension                                             &  quantity    & learnable      & description \\ \hline       
        $\mathbf{h}^{0\alpha}_{i}$                  & 4$N_a$                                                &  $N_e$       &                & one-electron features \\    
        $\mathbf{h}^{0\alpha\beta}_{ij}$            & 4                                                     &  $N_e^2$     &                & two-electron features \\    
        $\mathbf{h}^{\ell\alpha}_{i}$               & $n^{\ell-1}_1$                                        &  $(L-1) N_e$ &                & one-electron activations from layer $\ell-1$ \\
        $\mathbf{h}^{\ell\alpha\beta}_{ij}$         & $n^{\ell-1}_2$                                        &  $(L-1) N_e^2$ &              & two-electron activations from layer $\ell-1$ \\
        $\mathbf{f}^{\ell\alpha}_{i}$               & $3n^{\ell-1}_1 + 2n^{\ell-1}_2$                       &  $L N_e$     &                & one-electron input for layer $\ell$\\
        $\mathbf{V}^\ell$                           & $n^{\ell}_1\times(3n^{\ell-1}_1 + 2n^{\ell-1}_2)$     &  $L$         & \checkmark     & weights for one-electron linear layer \\
        $\mathbf{b}^\ell$                           & $n^{\ell}_1$                                          &  $L$         & \checkmark     & biases for one-electron linear layer \\
        $\mathbf{W}^\ell$                           & $n^{\ell}_2\times n^{\ell-1}_2$     &  $L$         & \checkmark     & weights for two-electron linear layer \\
        $\mathbf{c}^\ell$                           & $n^{\ell}_2$                                          &  $L$         & \checkmark     & biases for two-electron linear layer \\
        $\mathbf{w}^{k\alpha}_i $                   & $n^L_1$                                               &  $n_k N_e$   & \checkmark     & weights for final linear layer (orbital shaping) \\
        $g^{k\alpha}_i $                            & scalar                                                &  $n_k N_e$   & \checkmark     & bias for final linear layer (orbital shaping) \\
        $\pi_{im}^{k\alpha}$                        & scalar                                                &  $n_k N_a N_e$ & \checkmark     & enevelope weight \\
        $\Sigma_{im}^{k\alpha}$                     & $3 \times 3$                                            &  $n_k N_a N_e$ & \checkmark     & enevelope decay \\
        $\mathbf{\omega}$                           & $n_k$                                                 &  $1$         & \checkmark     & weights in determinant expansion \\
        \hline\hline
    \end{tabular}
    \caption{Network activations and parameters for Fermi-Net with $L$ layers, $n_k$ many-electron determinants for a system of $N_a$ atoms and $N_e$ electrons. $i,j$ index electrons in spin channels $\alpha,\beta \in \{\uparrow,\downarrow\}$. Each layer contains $n^{\ell}_1$ ($n^{\ell}_2$) hidden units for the one-electron (two-electron) stream. The quantity column shows the total number of each object.}
  \label{tab:parameters}
\end{table*}

\subsection{FermiNet architecture and training}

For all experiments, a Fermionic Neural Network with four layers was used, not counting the final linear layer that outputs the orbitals. Each layer had 256 hidden units for the one-electron stream and 32 hidden units for the two electron stream. A tanh nonlinearity was used for all layers, as a smooth function is needed to guarantee that the Laplacian is well defined and nonzero everywhere. 16 determinants were used where not otherwise specified. For comparison, the conventional VMC results in Table~\ref{tab:energy} from Seth {\em et al.} (2011)\cite{seth2011quantum} use 50 configuration state functions (CSF). While the exact number of determinants in a CSF will depend on the system, generally this will be on the order of hundreds to thousands of determinants. With this configuration of the FermiNet there were approximately 700,000 parameters in the network, although the exact number depends on the number of atoms in the system due to the way we construct the input features and exponentially-decaying envelope. A breakdown of these parameters are given in Table~\ref{tab:parameters}.

Before using the local energy as an optimization objective we pretrained the network to match Hartree-Fock (HF) orbitals computed using PySCF \cite{sun2018pyscf}. There were two reasons for this. First, we found that the numerical stability of the subsequent local energy optimization was improved. On large systems, the determinants in the Fermionic Neural Network would often numerically underflow if no pretraining was used, causing the optimization to fail. Pretraining with HF orbitals as a guide meant that the main optimization started in a region of relatively low variance, with comparitively stable determinant evaluations and electron walkers in representative configurations. Second, we found that time was saved by not optimizing the local energy through a region that we knew to be physically uninteresting, given that it had an energy higher than that of a straightforward mean field approximation. The pretraining did not seem to strand the neural network in a poor local optimum, as the energy minimization always gave consistent results capturing roughly 99\% of the correlation energy. This is consistent with the conventional wisdom in the machine learning community that issues with local minima are less severe in wider, deeper neural networks. Further, stochasticity in the optimization procedure helps break symmetry and escape bad minima.

The pretraining loss is:
\begin{equation*}
\begin{aligned}
	\mathcal{L}^{\mathrm{pre}}(\theta) = \int \left[\sum_{\alpha\in\{\uparrow, \downarrow\}} \sum_{ijk}\left(\phi^{k\alpha}_i\left(\mathbf{r}^{\alpha}_j; \{\mathbf{r}^\alpha_{/j}\}; \{\mathbf{r}^{\bar {\alpha}}\}\right)\right.\right. \\ - \left.\vphantom{\sum_{\alpha\in\{\uparrow, \downarrow\}}} \left. \phi^{\text{HF}}_{i\alpha}\left(\mathbf{r}^\alpha_j\right)\right)^2 \right] p^{\mathrm{pre}}(\mathbf{X}) d \mathbf{X} ,
\end{aligned}
\end{equation*}
where $\phi^{\text{HF}}_{i\alpha}\left(\mathbf{r}^\alpha_j\right)$ denotes the value of the $i$-th Hartree-Fock orbital for spin $\alpha$ at the position of electron $j$, $\bar{\alpha}$ is $\downarrow$ if $\alpha$ is $\uparrow$ or vice versa, and $\phi^{k\alpha}_i\left(\mathbf{r}^{\alpha}_j; \{\mathbf{r}^\alpha_{/j}\}; \{\mathbf{r}^{\bar{\alpha}}\}\right)$ is the corresponding entry in the input to the $k$-th determinant of the Fermionic Neural Network. We use a minimal (STO-3G) basis set for the Hartree-Fock computation as we require only a stable initialization in the rough vicinity of the mean field solution, not an accurate mean field result. The probability distribution $p^{\mathrm{pre}}(\mathbf{X})$ is an equal mixture of the product of Hartree-Fock orbitals and the output of the Fermionic Neural Network:
\begin{equation*}
p^{\mathrm{pre}}(\mathbf{X}) = \frac{1}{2}\left( \prod_{\alpha\in\{\uparrow, \downarrow\}}\prod_{i} (\phi^{\text{HF}}_{i\alpha}(\mathbf{r}^\alpha_i))^2 + \psi^2(\mathbf{X})\right) .
\end{equation*}
We chose not to use the distribution from the Hartree-Fock determinant because we wanted sample coverage at every point where the orbitals were large, but in practice the difference to using the anti-symmetrized distribution was marginal.  The inclusion of the neural network density helps to increase the sampling probability in areas where the neural network wavefunction is spuriously high. We approximate the expectation for the loss by using MCMC to draw half the samples in the batch from $\psi^2$ and half from the product of Hartree-Fock orbitals using MCMC.

Initial MCMC configurations were drawn from Gaussian distributions centred on each atom in the molecule. Electrons were assigned to atoms according to the nuclear charge and spin polarization of the ground state of the isolated atom, with the atomic spins orientated such that the total spin projection of the molecule was correct, which was possible for systems studied here. 
We used ADAM with default parameters as the optimizer. After pretraining, we reinitialized the electron walker positions and then had a burn in MCMC period with target distribution $\psi^2$ before we began local energy minimization.

For the FermiNet, all code was implemented in TensorFlow 1 built with CUDA 9. All experiments for systems with less than 20 electrons were run in parallel on 8 V100 GPUs, while 16 GPUs were used for larger systems. With a smaller batch size we were able to train on a single GPU but convergence was significantly and disproportionately slower. For instance, ethene converged after just 2 days of training with 8 GPUs, while several weeks were required on a single GPU. Bicyclobutane, with 30 electrons, took roughly 1 month on 16 GPUs to train. We expect further engineering improvements will reduce this number. 10 Metropolis-Hastings steps were taken between every parameter update, and it typically required $O(10^5-10^6)$ parameter updates to reach convergence (results in the paper used $2\times 10^5$ parameter updates). Conventional VMC wavefunction optimization will perform $O(10^1-10^2)$ parameter updates and $O(10^4-10^6)$ MCMC steps between updates, so we require roughly the same number of wavefunction evaluations as conventional VMC.  After network optimization, we run $O(10^5)$ MCMC steps and calculate the mean local energy every 10 steps. The energy and associated standard error are estimated using a standard approach to account for correlations.\cite{flyvbjerg1989}

\begin{table}[t]
    \centering
    \begin{tabular}{|c|c|c|}\hline
        & Parameter & Value \\\hline
       & Batch size & 4096 \\
       & Training iterations & 2e5 \\
       & Pretraining iterations & 1e3 \\
       & Learning rate & $(1e4 + t)^{-1}$ \\
       & Local energy clipping  & 5.0 \\
       KFAC & Momentum & 0 \\
       KFAC & Covariance moving average decay & 0.95 \\
       KFAC & Norm constraint & 1e-3 \\
       KFAC & Damping & 1e-3 \\
       MCMC & Proposal std dev (per dimension) & 0.02 \\
       MCMC & Steps between parameter updates & 10 \\
       \hline
    \end{tabular}
    \caption{Default hyperparameters for all experiments in the paper. For bicyclobutane, the batch size was halved and the pretraining iterations were increased by an order of magnitude.}
    \label{tab:hyperparams}
\end{table}

Accurate and stable convergence was highly dependent on the hyperparameters used; the default values for all experiments are included in Table~\ref{tab:hyperparams}. These hyperparameters do seem to be generalizable --- we have observed good performance on every system investigated. For some larger systems, stability was improved by using more pretraining iterations. Getting good performance from KFAC requires careful tuning, and we found that the damping and norm constraint parameters critically affect the asymptotic performance. If the damping is too high, KFAC behaves like gradient descent near a local minimum and converges too slowly. If the damping is reduced, training quickly becomes unstable unless the norm constraint (a generalization of gradient clipping) is lowered in tandem. Surprisingly, we found little advantage to using momentum, and sometimes it even seemed to reduce training performance, so we set it to zero for all experiments.

To reduce the variance in the parameter updates, we clipped the local energy when computing the gradients but not when evaluating the total energy of the system. This is a commonly used strategy to improve the accuracy of QMC\cite{umrigar1993diffusion}. We computed the total variation of each batch, $\frac{1}{N}\sum_i |E_L(\mathbf{X}_i)-\tilde{E}_L|$, where $\tilde{E}_L$ is the median local energy of that batch. This is to the $\ell_1$ norm what the standard deviation is to the $\ell_2$ norm, and we prefer it to the standard deviation as it is more robust to outliers. We clip any local energies more than 5 times further from the median than this total variation and compute the gradient in Eqn.~\ref{eqn:vmc_grad} with the clipped energy in place of $E_L$. The aforementioned KFAC norm constraint enforces gradient clipping in a manner which respects the information geometry of the model.

To sample from $\psi^2(\mathbf{X})$ we used the standard Metropolis-Hastings algorithm.\cite{foulkes2001quantum} The proposed moves were Gaussian distributed with a fixed, isotropic covariance. All electron positions were updated simultaneously. While one-electron moves are more common in VMC, prior work suggests that all-electron moves are effective at the scale of system we investigated,\cite{lee2011strategies} and the fact that our orbitals depend on all electrons means that we cannot exploit fast determinant updates with one-electron moves. We expect one-electron moves will have a more noticeable impact for larger systems and will investigate different MCMC strategies and parameters in future work. Typical acceptance rates were $\sim$0.95 for the smallest systems and $\sim$0.6 for the largest systems investigated. Due to slow equilibration of the MCMC sampling, the computed energy sometimes overshot the true value, but always reequilibrated after a few thousand iterations. We experimented with Hamiltonian Monte Carlo to give faster mixing and lower bias in the gradients, but found this led to significantly higher variance in the local energy and lower overall performance.

\subsection{Slater-Jastrow networks}
\label{sec:SJ_nets}

For the baseline Slater-Jastrow network, an multilayer perceptrons (MLP) with 3 hidden layers of 128 units were used for the orbitals. The electron positions and electron-nuclear vectors and distances were used as input features. The output of the MLP was fed into a final linear layer to generate the required orbitals and the same multiplicative envelope employed in the Fermionic Neural Network was included; this can be seen as an extension to the electron-nuclear Jastrow factor. The Jastrow factor and backflow transform are of the standard form:\cite{needs2015casino}
\begin{align}
    J(\{\mathbf{r}^\uparrow\}, \{\mathbf{r}^\downarrow\}, \{\mathbf{R}\}) = &J^{(e-n)}(\{\mathbf{r}^\uparrow\}, \{\mathbf{r}^\downarrow\}, \{\mathbf{R}\}) + \nonumber \\ &J^{(e-e)}(\{\mathbf{r}^\uparrow\}, \{\mathbf{r}^\downarrow\}) + \nonumber \\ 
    &J^{(e-e-n)}(\{\mathbf{r}^\uparrow\}, \{\mathbf{r}^\downarrow\}, \{\mathbf{R}\})
\end{align}
\begin{align}
    & J^{(e-n)}(\{\mathbf{r}^\uparrow\}, \{\mathbf{r}^\downarrow\}, \{\mathbf{R}\}) =\sum_{\alpha\in\{\uparrow, \downarrow\}} \sum_{i=1}^{n^\alpha} \sum_{I}^{N_a} \chi_j(|\mathbf{r}^{\alpha}_i - \mathbf{R}_I|) \nonumber \\
    &J^{(e-e)}(\{\mathbf{r}^\uparrow\}, \{\mathbf{r}^\downarrow\}) = \sum_{\alpha,\beta\in\{\uparrow, \downarrow\}} \sum_{i=1}^{n^\alpha} \sum_{j=1}^{n^\beta} u^{\alpha\beta}(|\mathbf{r}^{\alpha}_i - \mathbf{r}^{\beta}_j|) \nonumber \\
    & J^{(e-e-n)}(\{\mathbf{r}^\uparrow\}, \{\mathbf{r}^\downarrow\}, \{\mathbf{R}\}) = \sum_{\alpha,\beta\in\{\uparrow, \downarrow\}} \sum_{i=1}^{n^\alpha} \sum_{j=1}^{n^\beta}\sum_I^{N_a} \nonumber \\ &\hspace{1cm}f^{\alpha\beta}_k(|\mathbf{r}^{\alpha}_i - \mathbf{r}^{\beta}_j|, |\mathbf{r}^{\alpha}_i - \mathbf{R}_I|, |\mathbf{r}^{\beta}_j - \mathbf{R}_I|)
    \label{eqn:jastrow}
\end{align}
for the Jastrow factor and:
\begin{align}
    \mathbf{r}_i' = \mathbf{r}_i + &\mathbf{\xi}_i^{(e-e)}(\{\mathbf{r}_j\}) + \mathbf{\xi}_i^{(e-N)}(\{\mathbf{R}_I\}) \nonumber \\
    + &\mathbf{\xi}_i^{(e-e-N)}(\{\mathbf{r}_j\}, \{\mathbf{R}_I\})
\end{align}
\begin{align}
    &\mathbf{\xi}_i^{(e-e)}(\{\mathbf{r}_j\}) = \sum_{j\ne i}^n \eta(|\mathbf{r}_{ij}|) \mathbf{r}_{ij} \nonumber \\
    &\mathbf{\xi}_i^{(e-N)}(\{\mathbf{R}_I\}) = \sum_{I}^{N_a} \mu(|\mathbf{r}_{iI}|)\mathbf{r}_{iI} \nonumber \\
    &\mathbf{\xi}_i^{(e-e-N)}(\{\mathbf{r}_j\}, \{\mathbf{R}_I\}) = \nonumber\\
    &\sum_{j\ne i}^n \sum_{I}^{N_a} \Phi(|\mathbf{r}_{ij}|, |\mathbf{r}_{iI}|, |\mathbf{r}_{jI}|) \mathbf{r}_{ij} \nonumber \\
    &\hspace{0.8cm}+ \Theta(|\mathbf{r}_{ij}|, |\mathbf{r}_{iI}|, |\mathbf{r}_{jI}|)\mathbf{r}_{iI}
    \label{eqn:backflow}
\end{align}
for the backflow transform, where $\mathbf{r}_{ij}=\mathbf{r}_i-\mathbf{r}_j$ and $\mathbf{r}_{iI} = \mathbf{r}_i - \mathbf{R}_I$. Here $\{\chi_j\}$, $\{u^{\alpha\beta}\}$, $\{f^{\alpha\beta}_k\}$, $\eta$, $\mu$, $\Phi$ and $\Theta$ are all separate 3-layer perceptrons with 64 hidden units. Residual connections were used in all MLPs, which greatly improved the stability of training. We found Slater-Jastrow-backflow networks to be extremely unstable to train from random initial weights and hence used a fine-tuning approach where the Slater-Jastow-backflow networks were initialized from an optimized Slater-Jastrow network with the weights and biases in the backflow MLPs randomly initialized close to zero. The Slater-Jastrow and Slater-Jastrow-backflow networks were otherwise optimized in the identical fashion to FermiNet. 

\subsection{Hartree--Fock and Coupled Cluster calcuations}
\label{sec:HF_CC}

We used PySCF\cite{sun2018pyscf} to perform all-electron CCSD(T) calculations on atoms and dimers (Table~\ref{tab:energy}). PSI4\cite{parrish2017psi4} was used to perform all-electron CCSD(T) calculations on all other molecules, and and FCI calculations on H$_4$. Cholesky decomposition\cite{deprince2013cholesky} was used to reduce the memory requirements for bicyclobutane, which we verified introduces an error in the total energies of $\mathcal{O}(10^{-5})$ hartrees with the aug-cc-pCVTZ basis set. The H$_4$ calculations used a cc-pVXZ (X=T, Q, 5) basis set. All other CCSD(T) calculations used aug-cc-pCVXZ (X=T, Q, 5) basis sets. An unrestricted Hartree-Fock reference was used for atoms and dimers, with restricted Hartree-Fock used otherwise. We extrapolated energies to the CBS limit using standard methods\cite{feller1992application,helgaker1997basis}.
CBS Hartree-Fock energies for Li, Be and Li$_2$ were taken from aug-cc-pCV5Z calculations, in which the basis set error was below $10^{-4}$ hartrees.
CBS Hartree-Fock energies for other systems were obtained by fitting the function $E_{\mathrm{HF}}(X)=E_{\mathrm{HF}}(\mathrm{CBS})+ae^{-bX}$, where $X$ is the cardinality of the basis; 
CCSD, CCSD(T) and FCI correlation energies were extrapolated to the CBS by fitting the energies from quadruple- and quintuple-zeta basis sets (triple- and quadruple-zeta for bicyclobutane) to the function $E_c(X)=E_c(\mathrm{HF}) + aX^{-3}$.
The total energy is given by the sum of the Hartree-Fock energy and correlation energy. To compare the dissociation potential of N$_2$ against experiment, we used the MLR$_4$(6,~8) potential from Le Roy {\em et al.} (2006),\cite{leroy2006accurate} which is based on fitting 19 lines of the N$_2$ vibrational spectrum.

\section{Universality of Generalized Slater Determinants}
\label{sec:hutter}

Empirically, the accuracy of the FermiNet increases as the number of determinants grows. This raises the question: in theory, how many determinants are necessary to represent any antisymmetric function $\psi(\mathbf{x}_1,\ldots,\mathbf{x}_n)$ when the elements of the determinant are permutation-equivariant functions of the form $\mathbf{\Phi}_{ij} = \phi_{i}(\mathbf{x}_j; \{\mathbf{x}_{/j}\})$? The answer, perhaps surprisingly, is just one. The argument below is originally due to M. Hutter (personal communication).

Define a unique ordering on the vectors $\mathbf{x}_1,\ldots,\mathbf{x}_n$, for instance, $\mathbf{x}_i < \mathbf{x}_j$ if the first coordinate of $\mathbf{x}_i$ is less than the first coordinate of $\mathbf{x}_j$. Let $\pi$ be the permutation such that $\mathbf{x}_{\pi(1)} \le \mathbf{x}_{\pi(2)}\le\ldots\le \mathbf{x}_{\pi(n)}$, that is, $\pi$ sorts the vectors $\mathbf{x}_1,\ldots,\mathbf{x}_n$, and let $\sigma(\pi)$ be the sign of the permutation $\pi$. Define $\phi_1(\mathbf{x}_{j}; \{\mathbf{x}_{/j}\}) = \mathbbm{1}_{j=\pi(1)}\psi(\mathbf{x}_{\pi(1)},\ldots,\mathbf{x}_{\pi(n)})$ and $\phi_i(\mathbf{x}_{j}; \{\mathbf{x}_{/j}\}) = \mathbbm{1}_{j=\pi(i)}$ if $i\ne 1$. Then each row of the matrix has only one nonzero entry, and the determinant $\det[\mathbf{\Phi}_{ij}] = \sigma(\pi) \psi(\mathbf{x}_{\pi(1)},\ldots,\mathbf{x}_{\pi(n)}) = \psi(\mathbf{x}_1,\ldots,\mathbf{x}_n)$.

The functions $\phi_i$ are not everywhere continuous, due to the indicator functions $\mathbbm{1}_{j=\pi(i)}$, and therefore not learnable by the FermiNet. This may partially explain why, despite the theoretical universality of a single determinant, in practice we still require multiple determinants to achieve high accuracy. We should note that this construction is very similar to the suggestion in Luo and Clark\cite{luo2019backflow} that neural network backflow could be extended to continuous spaces by sorting the input vectors and multiplying a neural network Ansatz by the sign of the permutation. As the choice of ordering breaks a natural symmetry of the system, and the Ansatz becomes non-smooth anywhere the ordering changes, we suspect such an Ansatz would be less effective than the FermiNet, however it is appealingly simple.

\section{Equivalence of Natural Gradient Descent and Stochastic Reconfiguration}
\label{sec:ngd_and_sr}

Here we provide a derivation illustrating that stochastic reconfiguration is equivalent to natural gradient descent for unnormalized distributions. Though many authors have investigated extensions of the Fisher information metric to quantum systems, \cite{petz1996geometry} this particular connection between methods in machine learning and quantum chemistry seems not to be widely appreciated by either community, though it was pointed out in Nomura {\em et al.} (2017).\cite{nomura2017restricted}

We denote the density proportional to $\waveFunc^2(\overallState)$ by $p(\overallState)$, and the normalizing factor by $Z(\theta)$. In addition, let $\tilde{p}(\overallState)=\waveFunc^2(\overallState)$ denote the unnormalized density. In stochastic reconfiguration, the entries of the preconditioner matrix $\precond$ have the form
\[
\precond_{ij} = \expectp{\left(\mathcal{O}_i - \expectp{\mathcal{O}_i } \right)\left( \mathcal{O}_j - \expectp{\mathcal{O}_j} \right)},
\]
where
\[
\mathcal{O}_i(\overallState) = \waveFunc(\overallState)^{-1}\frac{\partial \waveFunc(\overallState)}{\partial \theta_i} = \frac{\partial \mathrm{log}|\waveFunc(\overallState)|}{\partial \theta_i} = \frac{1}{2}\frac{\partial \mathrm{log}\tilde{p}(\overallState)}{\partial \theta_i}
\]
and $\precond$ is a metric for the parameter space.\cite{mazzola_finite-temperature_2012} The term $\expectp{\mathcal{O}_i }$ can be expressed in terms of the normalizing factor:
\begin{align*}
\expectp{\mathcal{O}_i} &= \frac{1}{2}\expectp{\frac{\partial \mathrm{log}\tilde{p}(\overallState)}{\partial \theta_i} } \\
&= \frac{1}{2}\int \frac{\partial \mathrm{log}\tilde{p}(\overallState)}{\partial \theta_i} p(\overallState)d\overallState \\
&= \frac{1}{2}\int \frac{\partial \mathrm{log}\tilde{p}(\overallState)}{\partial \theta_i} \frac{\tilde{p}(\overallState)}{Z(\theta)}d\overallState \\
&= \frac{1}{2}\int \frac{1}{\tilde{p}(\overallState)}\frac{\partial \tilde{p}(\overallState)}{\partial \theta_i} \frac{\tilde{p}(\overallState)}{Z(\theta)}d\overallState \\
&= \frac{1}{2Z(\theta)}\int\frac{\partial \tilde{p}(\overallState)}{\partial \theta_i}d\overallState \\
&= \frac{1}{2Z(\theta)}\frac{\partial}{\partial \theta_i}\int \tilde{p}(\overallState)d\overallState \\
&= \frac{1}{2Z(\theta)}\frac{\partial Z(\theta)}{\partial \theta_i} \\
&= \frac{1}{2}\frac{\partial \mathrm{log}Z(\theta)}{\partial \theta_i}.
\end{align*}
Plugging this into the expression for $\precond_{ij}$ yields
\begin{align*}
\precond_{ij} =& \expectp{\left(\mathcal{O}_i - \expectp{\mathcal{O}_i } \right)\left( \mathcal{O}_j - \expectp{\mathcal{O}_j} \right)} \\
	=& \frac{1}{4} \mathbb{E}_{p(\overallState)}
		\begin{aligned}[t]
			&\left[\left(\frac{\partial \mathrm{log}\tilde{p}(\overallState)}{\partial \theta_i} -\frac{\partial \mathrm{log}Z(\theta)}{\partial \theta_i} \right)\right. \\
			&\hphantom{\left[\right.}\left.\left( \frac{\partial \mathrm{log}\tilde{p}(\overallState)}{\partial \theta_j}- \frac{\partial \mathrm{log}Z(\theta)}{\partial \theta_j} \right)\right]
		\end{aligned}
	\\
=& \frac{1}{4}\expectp{\frac{\partial \mathrm{log}p(\overallState)}{\partial \theta_i}  \frac{\partial \mathrm{log}p(\overallState)}{\partial \theta_j}},
\end{align*}
which, up to a constant, is the Fisher information metric for $p(\overallState)$.

\section{Numerically Stable Computation of the Log Determinant and Derivatives}
\label{sec:numerical_stability}
For numerical stability, the Fermionic Neural Network outputs the {\em logarithm} of the absolute value of the wavefunction (along with its sign), and we compute log determinants rather than determinants. Even if some of the matrices are singular, this is not an issue for numerical stability on the forward pass, because these matrices will have zero contribution to the overall sum of determinants the network outputs:
\[
\mathrm{log}|\psi(\mathbf{r}^\uparrow_1,\ldots,\mathbf{r}^\downarrow_{n^\downarrow})| = \mathrm{log} \left|\sum_{k}\omega_k \det\left[\mathbf{\Phi}^{k \uparrow}\right]\det\left[\mathbf{\Phi}^{k\downarrow}\right]\right|.
\]
We use the ``log-sum-exp trick" to compute the sum --- that is, we subtract off the largest log determinant before exponentiating and computing the weighted sum, and add it back in after the logarithm at the end. This avoids numerical underflow if the log determinants are not well scaled.

Naively applying automatic differentiation frameworks to compute the gradient and Laplacian of the log wavefunction will not work if one of the matrices is singular. However, the first and second derivatives are still well defined, and we show how to express these derivatives in closed form appropriate for reverse-mode automatic differentiation. Several of the results used here, as well as the notation, are based on the collected matrix derivative results of Giles (2008)\cite{giles2008collected}.

From Jacobi's formula, the gradient of the determinant of a matrix is given by
\[
\frac{\partial \det(\genericMatrix)}{\partial \genericMatrix} = \det(\genericMatrix)\genericMatrix^{-T}=\mathrm{Adj}(\genericMatrix)^T = \mathrm{Cof}(\genericMatrix),
\]
where $\mathrm{Cof}(\genericMatrix)$ is the cofactor matrix of $\genericMatrix$. Let $\cofMat = \mathrm{Cof}(\genericMatrix)$. Then, by the product rule, we can express the reverse-mode gradient of $\mathrm{Cof}(\genericMatrix)$ as
\[
\bar{\genericMatrix} = \genericMatrix^{-T}\left[\mathrm{Tr}\left(\bar{\cofMat}^T\mathrm{Cof}(\genericMatrix)\right)\identity-\bar{\cofMat}^T\mathrm{Cof}(\genericMatrix)\right],
\]
where $\bar{\cofMat}$ is the reverse-mode sensitivity. Unfortunately, this expression becomes undefined if the matrix $\genericMatrix$ is singular. Even so, both the cofactor matrix and its derivative are still well defined. To see this, we express the cofactor in terms of the singular value decomposition of $\genericMatrix$. Let $\svdLeft\svdMiddle\svdRight^T$ be the singular value decomposition of $\genericMatrix$, then
\begin{align*}
\mathrm{Cof}(\genericMatrix) &= \det(\genericMatrix)\genericMatrix^{-T} \\
 &= \det(\svdLeft)\det(\svdMiddle)\det(\svdRight)\svdLeft\svdMiddle^{-1}\svdRight^T .
\end{align*}
Since $\svdLeft$ and $\svdRight$ are orthonormal matrices, their determinant is just the sign of their determinant. To avoid clutter, we drop the $\det(\svdLeft)$ and $\det(\svdRight)$ terms until the very end. Let $\sigma_i$ be the $i$th diagonal element of $\svdMiddle$, then we have $\det(\svdMiddle)=\prod_i \sigma_i$, and cancelling terms in the expression, we get (up to a sign factor)
\[
\mathrm{Cof}(\genericMatrix) = \svdLeft\mathbf{\Gamma} \svdRight^T,
\]
where $\mathbf{\Gamma}$ is a diagonal matrix with elements $\gamma_i$ defined as
\[
\gamma_i = \prod_{j\ne i} \sigma_j
\]
because the $\sigma_i^{-1}$ term in $\svdMiddle^{-1}$ cancels out one term in $\det(\svdMiddle)$.


\begin{table}
    \centering
    \addtolength{\tabcolsep}{3pt}
    \begin{tabular}{rll}\hline\hline
    $N$ &   \multicolumn{2}{c}{Energy / $N$ ($E_h$)} \\\cline{2-3}
    & Separation: 10 $a_0$ & Separation: 15 $a_0$ \\ \hline
             2 &   -0.5000023(9) & -0.50000021(5) \\
             4 &   -0.4999977(6) & -0.4999991(2) \\
             6 &   -0.499991(2) & -0.499990(1) \\
             8 &   -0.499985(3) & -0.499993(2) \\
            10 &   -0.499980(2) & -0.499989(1) \\
    \hline\hline
    \end{tabular}
    \addtolength{\tabcolsep}{-3pt}
    \caption{Chains of $N$ hydrogen atoms at equal separations. The energy per atom is in excellent agreement with that of a single hydrogen atom.}
    \label{tab:hn_sep}
\end{table}

The gradient of the cofactor is more complicated, but once again terms cancel. Again neglecting a sign factor, the reverse-mode gradient can be expanded in terms of the singular vectors as:
\begin{align*}
\bar{\genericMatrix} &= \genericMatrix^{-T}\left[\mathrm{Tr}\left(\bar{\cofMat}^T\mathrm{Cof}(\genericMatrix)\right)\identity-\bar{\cofMat}^T\mathrm{Cof}(\genericMatrix)\right] \\
&= \svdLeft\svdMiddle^{-1}\svdRight^T\left[\mathrm{Tr}\left(\bar{\cofMat}^T \svdLeft\mathbf{\Gamma}\svdRight^T\right)\identity-\bar{\cofMat}^T \svdLeft\mathbf{\Gamma} \svdRight^T\right] \\
&= \svdLeft\left[\mathrm{Tr}\left(\bar{\cofMat}^T \svdLeft\mathbf{\Gamma} \svdRight^T\right)\svdMiddle^{-1}-\svdMiddle^{-1}\svdRight^T\bar{\cofMat}^T \svdLeft\mathbf{\Gamma}\right]\svdRight^T \\
&= \svdLeft\left[\mathrm{Tr}\left(\mathbf{M}\mathbf{\Gamma}\right)\svdMiddle^{-1}-\svdMiddle^{-1}\mathbf{M}\mathbf{\Gamma}\right]\svdRight^T,
\end{align*}
where $\mathbf{M}=\svdRight^T\bar{\cofMat}^T \svdLeft$, and we have taken advantage of the invariance of the trace of matrix products to cyclic permutation in the last line.

Now, in the expression inside the square brackets in the last line, terms conveniently cancel that prevent the expression from becoming undefined should $\sigma_i=0$ for some singular value. Denote this term $\mathbf{\Xi}$, the off-diagonal terms of $\mathbf{\Xi}$ only depend on the second term $\svdMiddle^{-1}\mathbf{M}\mathbf{\Gamma}$:
\begin{align*}
\Xi_{ij} &= -M_{ij}\sigma_i^{-1}\gamma_j \\
&= -M_{ij}\sigma_i^{-1}\prod_{k \ne j} \sigma_k \\
&= -M_{ij}\prod_{k \ne i,j} \sigma_k ,
\end{align*}
and the diagonal terms have the form
\begin{align*}
\Xi_{ii} &= \sigma_i^{-1} \sum_{j} M_{jj}\gamma_j - M_{ii}\sigma_i^{-1}\gamma_i \\
&= \sum_{j \ne i} M_{jj}\sigma_i^{-1}\gamma_j \\
&= \sum_{j \ne i} M_{jj}\prod_{k \ne i,j} \sigma_k .
\end{align*}
Putting this all together, we get
\[
\bar{\genericMatrix} = \mathrm{Sgn}(\det(\svdLeft))\mathrm{Sgn}(\det(\svdRight))\svdLeft\mathbf{\Xi} \svdRight^T ,
\]
with
\[
\Xi_{ij} =
    \begin{cases}
      \sum_{j \ne i} M_{jj} \rho_{ij}, & \text{if}\ i=j, \\
      -M_{ij}\rho_{ij}, & \text{otherwise,}
      \end{cases}
\]
\[
\rho_{ij} = \prod_{k \ne i,j} \sigma_k,
\]
\[
\mathbf{M} = \svdRight^T\bar{\cofMat}^T\svdLeft.
\]
This allows us to compute second derivatives of the matrix determinant even for singular matrices. To handle degenerate matrices gracefully, we fuse everything from the computation of the log determinant to the final network output into a single TensorFlow operation, with a custom gradient and gradient-of-gradient that includes the expression above.

\begin{figure}
    \centering
    \includegraphics[width=\columnwidth]{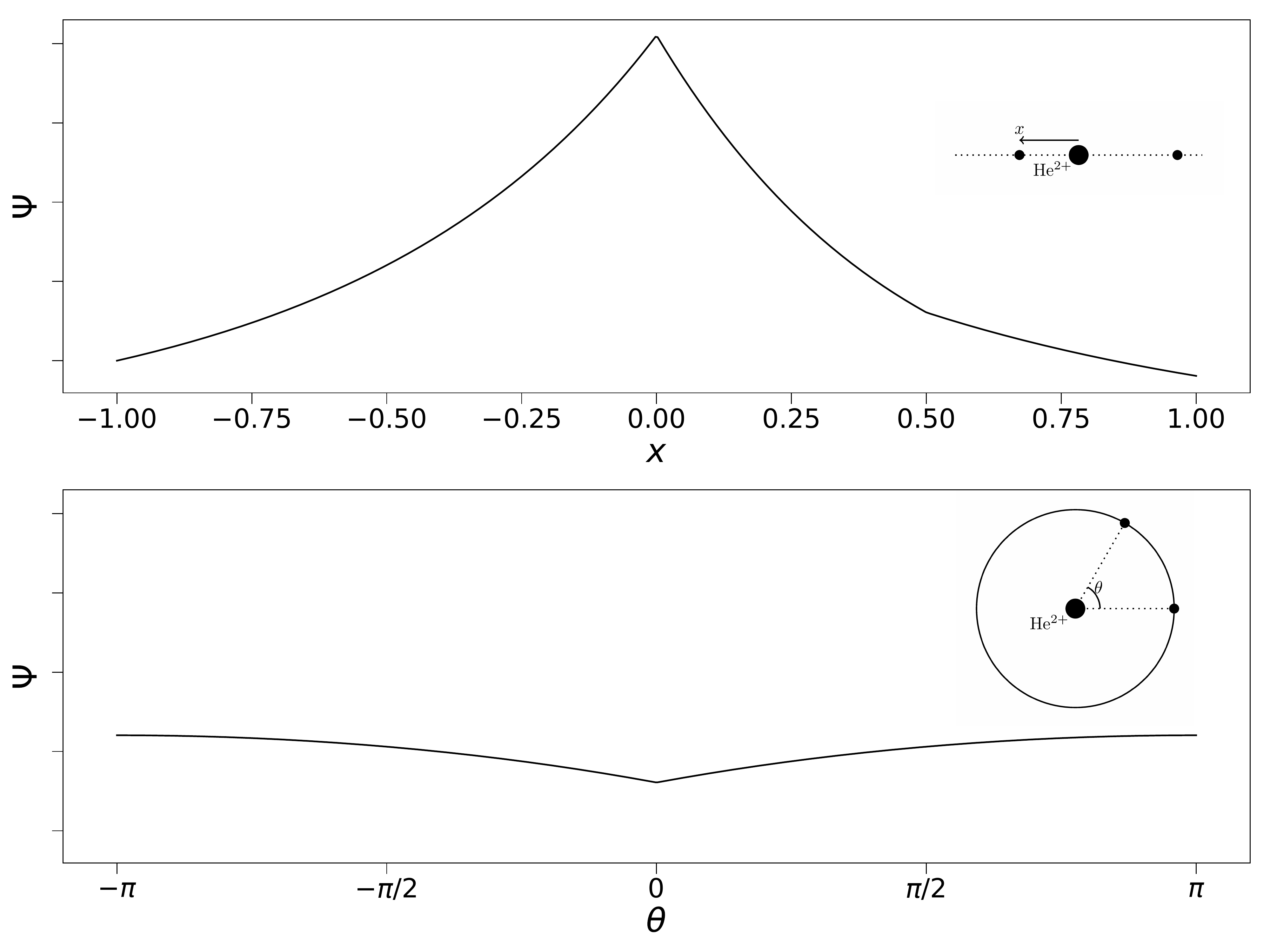}
    \caption{Evaluation of the FermiNet wavefunction for the helium atom. The second electron is clamped at position $(0.5,0,0) a_0$ and the first electron is moved along the path $(x,0,0)a_0$, through both the nucleus and the second electron (top), and along the path $(0.5\cos\theta, 0.5\sin\theta,0)a_0$, through the second electron (bottom).}
    \label{fig:he_cusps}
\end{figure}

\section{Non-interacting hydrogen chains}
\label{sec:hn_consistency}

At sufficently large separations, two systems become non-interacting. The energy of the combined system should be equal to the sum of the energies of the individual systems. We demonstrate this property for FermiNet on chains of well-separated hydrogen atoms of up to 10 atoms (Table~\ref{tab:hn_sep}).

\section{Electron-Electron and Electron-Nuclear Cusps}
\label{sec:cusps}

The derivatives of the wavefunction must be discontinuous when two electrons or an electron and nucleus coincide in order to cancel corresponding singularities in the Hamiltonian. Capturing these cusps correctly, especially the electron-nuclear cusp, is critical for accurately capturing correlation energy. Assuming the wavefunction is non-zero at these points, the cusp conditions specify the relationship between the wavefunction and its derivative to be:
\begin{align*}
\lim_{r_{iI}\rightarrow 0}\left(\frac{\partial\Psi}{\partial r_{iI}}\right)_{\textrm{ave}} &= -Z\Psi(r_{iI} = 0) \\
\lim_{r_{ij}\rightarrow 0}\left(\frac{\partial\Psi}{\partial r_{ij}}\right)_{\textrm{ave}} &= \frac{1}{2} \Psi(r_{ij} = 0)
\end{align*}
where $r_{iI}$ ($r_{ij}$) is an electron-nuclear (electron-electron) distance, $Z_I$ is the nuclear charge of the $I$-th nucleus and $\textrm{ave}$ implies a spherical averaging over all directions.

Fig.~\ref{fig:he_cusps} shows FermiNet correctly describes the cusps for the helium atom. We estimate $\lim_{r\rightarrow 0}\frac{\partial\log|\Psi|}{\partial r}$ using Monte Carlo integration over spherical surfaces of radius $10^{-5}a_0$ centered on the helium nucleus and second electron, fixed at $0.5a_0$ from the nucleus, and obtain, where $r_1$ ($r_{12}$) is the distance between the first electron and the nucleus (second electron), 
\begin{align*}
\left(\frac{\partial\log|\Psi|}{dr_1}\right)_{r_1=0, \textrm{ave}} &= -1.9979(4) \\
\left(\frac{\partial\log|\Psi|}{dr_{12}}\right)_{r_{12}=0, \textrm{ave}} &= 0.4934(1),
\end{align*}
in excellent agreement with the theoretical values.

\section{Molecular structures}
\label{sec:geometries}
Molecular structures were taken from the G3 database\cite{curtiss1998gaussian} where available. We reproduce the atomic positions for all molecules studied in Tables~\ref{tab:nh3_structure}-\ref{tab:bicyclobutane_structure}.

\begin{table}[H]
    \centering
    \begin{tabular}{|c|c|}\hline
    Atom & Position ($a_0$) \\\hline
    N & (0.0, 0.0, 0.22013) \\
    H1 & (0.0, 1.77583, -0.51364) \\
    H2 & (1.53791, -0.88791, -0.51364) \\
    H3 & (-1.53791, -0.88791, -0.51364) \\\hline
    \end{tabular}
    \caption{Atomic positions for ammonia (NH$_3$).}
    \label{tab:nh3_structure}
\end{table}
\medskip
\medskip

\begin{table}[H]
    \centering
    \begin{tabular}{|c|c|}\hline
    Atom & Position ($a_0$) \\\hline
    C & (0.0, 0.0, 0.0) \\
    H1 & (1.18886, 1.18886, 1.18886) \\
    H2 & (-1.18886, -1.18886, 1.18886) \\
    H3 & (1.18886, -1.18886, -1.18886) \\
    H4 & (-1.18886, 1.18886, -1.18886) \\\hline
    \end{tabular}
    \caption{Atomic positions for methane (CH$_4$).}
    \label{tab:ch4_structure}
\end{table}

\begin{table}[H]
    \centering
    \begin{tabular}{|c|c|}\hline
    Atom & Position ($a_0$) \\\hline
    C1 & (0.0, 0.0, 1.26135) \\
    C2 & (0.0, 0.0, -1.26135) \\
    H1 & (0.0, 1.74390, 2.33889) \\
    H2 & (0.0, -1.74390, 2.33889) \\
    H3 & (0.0, 1.74390, -2.33889) \\
    H4 & (0.0, -1.74390, -2.33889) \\\hline
    \end{tabular}
    \caption{Atomic positions for ethene (C$_2$H$_4$).}
    \label{tab:ethene_structure}
\end{table}

\begin{table}[H]
    \centering
    \begin{tabular}{|c|c|}\hline
    Atom & Position ($a_0$) \\\hline
    C & (0.0517, 0.7044, 0.0) \\
    N & (0.0517, -0.7596, 0.0) \\
    H1 & (-0.9417, 1.1762, 0.0) \\
    H2 & (-0.4582, -1.0994, 0.8124) \\
    H3 & (-0.4582, -1.0994, -0.8124) \\
    H4 & (0.5928, 1.0567, 0.8807) \\
    H5 & (0.5928, 1.0567, -0.8807) \\
    \hline
    \end{tabular}
    \caption{Atomic positions for methylamine (CH$_3$NH$_2$).}
    \label{tab:methylamine_structure}
\end{table}

\begin{table}[H]
    \centering
    \begin{tabular}{|c|c|}\hline
    Atom & Position ($a_0$) \\\hline
    \hline
    O1 & (0.0, 2.0859, -0.4319) \\
    O2 & (0.0, 0.0, 0.8638)  \\
    O3 & (0.0, -2.0859, -0.4319) \\
    \hline
    \end{tabular}
    \caption{Atomic positions for ozone (O$_3$).}
    \label{tab:ozone_structure}
\end{table}

\begin{table}[H]
    \centering
    \begin{tabular}{|c|c|}\hline
    Atom & Position ($a_0$) \\\hline
    C1 & (2.2075, -0.7566, 0.0) \\
    C2 & (0.0, 1.0572, 0.0) \\
    O  & (-2.2489, -0.4302, 0.0) \\
    H1 & (-3.6786, 0.7210, 0.0) \\
    H2 & (0.0804, 2.2819, 1.6761) \\
    H3 & (0.0804, 2.2819, -1.6761) \\
    H4 & (3.9985, 0.2736, 0.0) \\
    H5 & (2.1327, -1.9601, 1.6741) \\
    \hline
    \end{tabular}
    \caption{Atomic positions for ethanol (C$_2$H$_5$OH).}
    \label{tab:ethanol_structure}
\end{table}

\begin{table}[H]
    \centering
    \begin{tabular}{|c|c|}\hline
    Atom & Position ($a_0$) \\\hline
    C1 & (0.0, 2.13792, 0.58661) \\
    C2 & (0.0, -2.13792, 0.58661) \\
    C3 & (1.41342, 0.0, -0.58924) \\
    C4 & (-1.41342, 0.0, -0.58924) \\
    H1 & (0.0, 2.33765, 2.64110) \\
    H2 & (0.0, 3.92566, -0.43023) \\
    H3 & (0.0, -2.33765, 2.64110) \\
    H4 & (0.0, -3.92566, -0.43023) \\
    H5 & (2.67285, 0.0, -2.19514) \\
    H6 & (-2.67285, 0.0, -2.19514) \\\hline
    \end{tabular}
    \caption{Atomic positions for bicyclobutane (C$_4$H$_6$).}
    \label{tab:bicyclobutane_structure}
\end{table}
\vfill\null

\bibliography{references}

\end{document}

%% file: preamble.tex
\usepackage{amsmath}
\usepackage{amsfonts}
\usepackage[utf8]{inputenc}
\usepackage{xcolor}
\usepackage{bbm}
\usepackage{graphicx}
\usepackage{float}
\usepackage{hhline}
\usepackage{booktabs}
\usepackage{etoolbox}
\usepackage{bbm}
\usepackage{cleveref}
\restylefloat{table}

\usepackage{tikz}
\usetikzlibrary{shapes,arrows}
\input{tikz_defs.tikz}

\newcommand{\commentOut}[1]{}

\newcommand{\resolved}[1]{}

\newcommand{\waveFunc}{\psi}

\newcommand{\overallState}{\mathbf{X}}

\newcommand{\localEnergy}{{E_L}}

\newcommand{\qmOperator}[1]{\hat{#1}}
\newcommand{\hamiltonian}{\qmOperator{H}}

\newcommand{\genericMatrix}{\mathbf{A}}

\newcommand{\precond}{\mathcal{M}}

\newcommand{\identity}{\mathbf{I}}
\newcommand{\cofMat}{\mathbf{C}}
\newcommand{\svdLeft}{\mathbf{U}}
\newcommand{\svdMiddle}{\mathbf{\Sigma}}
\newcommand{\svdRight}{\mathbf{V}}

\newcommand{\expect}[2]{\mathbb{E}_{#1}\left[#2\right]}
\newcommand{\expectp}[1]{\expect{p(\overallState)}{#1}}

\newtoggle{arxivversion} 

%% file: tikz_defs.tikz
\tikzstyle{start} = [rectangle, draw, fill=red!20, 
    text width=5.5em, text centered, rounded corners, minimum height=4em]
\tikzstyle{block} = [rectangle, draw, fill=blue!20, 
    text width=5.5em, text centered, rounded corners, minimum height=4em]
\tikzstyle{terminal} = [rectangle, draw, fill=yellow!20, 
    text width=5.5em, text centered, rounded corners, minimum height=4em]
\tikzstyle{line} = [draw, -latex']